\documentclass[twocolumn,secnumarabic,amssymb, nobibnotes, aps, prappl, floatfix,longbibliography]{revtex4-2}

\usepackage{graphicx}
\usepackage[section]{placeins}
\usepackage{natbib}
\usepackage{xcolor}
\usepackage{comment}
\usepackage{hyperref}
\usepackage{upgreek}

\newcommand{\degC}{$^{\circ}$C}
\hypersetup{
    colorlinks=true,        
    linkcolor=blue,          
    citecolor=blue,         
    urlcolor=blue           
}

\begin{document}

\title{In-situ Investigation of the Phase Formation and Superconductivity in V$_3$Si Thin Films at High Temperatures}%

\author{$^1$M. Bose, $^2$D.L. Cortie, $^3$S. Rubanov, $^2$A.P. Le Brun,$^1$T.R. Finlayson and $^1$J.C. McCallum}%
\affiliation{$^1$School of Physics, The University of Melbourne, Parkville, Victoria 3010\\
$^2$Australian Nuclear Science and Technology Organisation, NSW, Australia\\
$^3$Ian Holmes Imaging Centre, Bio21 Institute, University of Melbourne, Parkville, Victoria 3010
}

\begin{abstract}
Vanadium silicide (V$_3$Si) is a promising superconductor for integration with silicon-based  electronics, however the interfacial growth kinetics  have a strong influence on the resulting superconducting properties and are not yet fully understood. In this study, we have used neutron reflectometry to reveal the phase transformation  during thin film growth driven by different annealing strategies. We examined the silicide formation when a thin layer of vanadium undergoes reactive diffusion  with a silicon dioxide film on silicon at temperatures from 650--800 \degC. To further investigate the time evolution of different phases under various annealing temperatures, a chemical model was developed and subsequent simulations were performed. The results of this model were validated using X-ray diffraction and cross-sectional TEM analysis. Correlations were observed between the structure and superconducting properties. Over-annealing films leads to complete depletion of the SiO$_2$ barrier layer, forming diffuse interfaces and driving the formation of undesirable silicon-rich  silicides.  Avoiding this by controlling time and temperature, allows higher quality superconducting films to be achieved. The $T_c$ of the films was found to be 13 K, and the annealing conditions influenced the critical fields and the paramagnetic Meissner effect near $T_c$. For optimally--annealed films, superconducting order parameters were calculated. Ginzberg-Landau theory was applied to explain flux penetration.
\end{abstract}

\maketitle

\section{Introduction}

Superconducting quantum circuits have gained prominence in various quantum technologies. They form part of mainstream quantum computing applications due to their ease of integration with traditional Complementary Metal-Oxide-Semiconductor (CMOS) technology and their very short gating times in the order of nanoseconds \citep{kjaergaard2020}.  Incorporation of both superconducting and semiconducting components within a single semiconductor device has great potential in the development of novel quantum device structures \citep{burkard2020}. However, increasing scale and fidelity are still bottlenecks in the development of superconducting quantum computers, especially with large scale industry demands. Also, the current superconducting qubits need to be cooled down to cryogenic temperatures, usually at millikelvin, using a dilution refrigerator, for coherent manipulation of quantum states \citep{bravyi2022}. If higher qubit operating temperatures can be realized, this can increase the available cooling power, leading to improved scalability and integration of qubits \citep{yang2020,petit2020}. In this context, choosing a superconducting material with a high critical temperature that can be easily integrated with classical electronics is pertinent.

The A15 family of superconductors, which exhibit the $\beta$-W crystal structure with the cubic space group Pm3, have been well-known for the past 70 years. Among these, superconductivity in the binary compound vanadium silicide (V$_3$Si) having a superconducting critical temperature, $\textit{T}_{\text{c}}$, of 17.1 K, was discovered by Hardy G.F. and Hulm J.K. \citep{hardy1953, hardy1954}. The realization of an upper critical field in excess of 23 T \cite{Saur1969} resulted in intensive research on this material for potential applications. The successful growth of single crystals of V$_3$Si \citep{Greiner1964} resulted in the subsequent discovery of a structural transformation at a temperature, $\textit{T}_{\text{m}}$, just a few degrees above $\textit{T}_{\text{c}}$, by both neutron \citep{Shull1966} and x-ray diffraction \citep{Batterman1966}, as well as by other techniques such as the temperature dependence of the elastic constant, ($C_{11}$ - $C_{12}$)/2 \cite{Testardi1967}, and specific heat \citep{Kunzler1966}. In the case of the elastic constant measurements, while all single crystals studied exhibited intensive elastic softening with temperature from room temperature to just above $\textit{T}_{\text{c}}$, not all samples exhibited the structural transformation \citep{Testardi1967}. X-ray diffraction studies of polycrystalline V$_3$Si have also revealed the structural transformation a few degrees above $\textit{T}_{\text{c}}$ in some samples \citep{King1967}, Consequently, there has been a wealth of research both theoretical \citep{mattheiss1982,bisi1982} and experimental \citep{Testardi1975, Dierker1983, Mitrovi1986, Hirota1995, Suga2012}, attempting to explain the relationship between the structural behaviour and superconductivity. More recently, the superconductivity of V$_3$Si thin films has received attention in the literature \citep{Howard2023,zhang2021}, on account of the potential of a thin film of V$_3$Si grown on a suitable substrate, for device development.

V$_3$Si thin films can be developed through various methods, including sputtering \citep{Bangert1985}, dual-beam evaporation of vanadium and silicon \citep{Borghesi1989}, and e-beam evaporation \citep{Oya1982,Psaras1984} of vanadium on different substrates. In most cases, a subsequent high-temperature annealing step is necessary for silicide formation with the desired stoichiometry.
A15 superconducting thin films are receiving renewed attention in recent times due to their potential use in spintronics, primarily due to their manifestation of the spin Hall effect (SHE) \citep{derunova2019} and in quantum computing applications \citep{vethaak2021}, driven by their high critical temperature and critical field. Therefore, understanding the phase transformation processes involved in the formation of the V$_3$Si films and their superconducting properties are of importance. However, there are difficulties associated with developing thin films, primarily in controlling the thickness of the V$_3$Si layer. The growth kinetics such as annealing temperature and time that directly affect the film's thickness, interface roughnesses and superconducting quality are not yet completely understood. Hence, it is of importance to determine the optimal growth parameters that help to achieve precise control over the superconducting layer. The chemical mechanisms that govern the growth kinetics and their effect on the layer properties are also worthy of attention to identify how varying growth conditions modify the superconducting properties of the films. Addressing these challenges is critical in advancing the use of V$_3$Si in quantum devices, where stable and reproducible device outcomes and performance will be required across different advanced fabrication technologies. 

In this work, we performed in-situ annealing of the prepared V thin film samples and investigated the phase changes at different temperatures using neutron reflectometry for the first time. Neutron reflectometry is a technique that can be used to understand the surface and interface structure down to the nanoscale (0.5-300 nm thickness range). Neutrons are highly penetrating, making them the best possible source for in-situ reflectometry studies \citep{imae2011neutrons}. V$_3$Si single crystals have previously been subjected to neutron scattering experiments for studies of acoustic-phonon softening \citep{Shirane1971}, phonon density of states \citep{knapp1976} and the flux-line lattice (FLL) transition \citep{yethiraj2005}.

Regardless of the deposition method for the V$_3$Si films, the major challenge is achieving the correct 3:1 phase ratio. This is because there are several factors affecting the phase transformation, such as ratio between V and Si, annealing conditions such as temperature, time and the presence of oxygen. The neutron reflectometry method proposed in this study is capable of addressing these challenges by revealing evolution of the phase changes during the in-situ annealing. For this study we use the e-beam method, where metallic V thin films are evaporated on to an oxidized silicon substrate. The e-beam evaporation is a cost-effective way of developing thin films and hence it is of interest to identify the phase transformations during formation of V$_3$Si by this pathway.

After the phase transformation studies, the superconducting properties of the samples were analyzed. The second half of this paper is completely dedicated to the discussion of magnetometry measurement results, that include details of the superconducting order parameters. We also examine the flux penetration in to these films, based on vortex core size using Ginzburg-Landau theory.

\section{Experimental Technique}

\subsection{Sample Preparation}

In this study, we utilized p-type silicon substrates measuring 2 x 2 cm$^2$ with a room temperature resistivity of 10 ohm-cm. The sample preparation involved the thermal growth of a 200 nm silicon dioxide (SiO$_2$) layer on the silicon substrate through dry oxidation at 1000$^\circ$C. Subsequently, a 300 nm thick V layer was deposited by electron-beam evaporation at a rate of 7 \AA/s, with a base pressure of 2 × 10$^{-7}$ mbar.

\subsection{Neutron Reflectometry Method}

Neutron reflection involves the specular reflection of neutron planewaves from a 1D surface potential, which can be used to profile the isotopic profile of thin films and surfaces with a depth resolution between 1-300 nm. The process can be described by a 1D time-independent Schrodinger equation:
\begin{equation}
    \left[ -\frac{\hbar^2}{2m_n}\frac{\delta}{\delta z^2} + V(z) \right] \Psi(z) = E\Psi(z).
    \label{eq:schrodinger}
\end{equation}
where for unpolarized neutrons the potential is:
\begin{equation}
  V(z)= \frac{2\pi\hbar^2}{m_n} \rho_{N} (z) , 
  \label{eq:pnr_potential}
\end{equation}
where $m_n$ is the neutron mass, $\rho_{N}$ is the nuclear scattering length density (SLD):
\begin{equation}
    \rho_N= \sum_{i}^{}N_i b_i,
\end{equation}
where the sum runs over the different isotopes in each layer, each with number density $N_i$ and neutron coherent scattering length $b_i$. The numerical methods to model the process are well known, and ultimately boundary conditions at different interfaces determine the neutron wave function, and the amplitude of the neutron wave function $|\Psi|^2$ that is measured at the detector. In practice, the mathematics of neutron reflection is often understood using analogous optical wave equations, such as those that apply to X-ray reflectometry.

Neutron reflectometry measurements were conducted using the Spatz neutron beam instrument at the 20 MW OPAL research reactor (Australian Nuclear Science and Technology Organisation (ANSTO), Lucas Heights, Australia) \citep{le2023spatz}. The Spatz instrument views the cold neutron source and uses the time-of-flight (ToF) principle. The disc chopper settings used were a pairing of discs 1 and 2 set 480 mm apart for a fractional wave length resolution $\frac{\delta\lambda}{\lambda}$ of 5\%. The chopper discs were rotating at a speed of 25 Hz to generate a usable neutron bandwidth of wavelengths 2.8 to 18 ${\textup{\,\AA}}$. The samples were illuminated with a 20 mm footprint along the beam and used a first angle of incidence of 0.6$^{\circ}$ with collimation slit settings of 1.90 mm and 0.14 mm and a second angle of incidence of 2.7$^{\circ}$ with slit settings of 8.56 mm and 0.62 mm. Counting time for each angle was 2400 s and 5400 s, respectively. 

Collected data were reduced within the \textit{refnx} package \citep{nelson2019refnx} where the reflectometry data are normalized to a direct beam  measurement that is taken over the same collimation conditions. ToF is converted to wavelength by using the de Broglie equation,  corrected for detector efficiency, joining the two data sets for each angle of incidence at an appropriate overlap region, and scaling the data so that the critical edge is equal to a reflectivity of one. The final data are recorded as momentum transfer, $Q$ (in units of  ${\textup{\,\AA}}^{-1}$), reflectivity, $R$, its standard deviation, $\mathrm{d}R$, and the instrument resolution, $\mathrm{d}Q$. The momentum transfer, $Q$, which is the scattering vector, is calculated as, 
\begin{equation}
Q=\frac{(4\pi sin\theta)}{\lambda},
\end{equation}
where  $\theta$ is the angle of incidence and $\lambda$ is the neutron wavelength. 
All data were collected in `event mode' where each neutron has its position on the detector and its time recorded, allowing data to be reduced to desired time intervals post-data collection. Data for the 0.6$^{\circ}$ angle of incidence were additionally reduced as described above, but also at 600 s time intervals, over the 2400 s to determine any time dependent changes over a given temperature.    

Data were also fitted within the \textit{refnx} software package \citep{nelson2019refnx}. In the models used the samples are divided into a series of layers, defined in terms of thickness (${\textup{\,\AA}}$), roughness (in units of $\textup{\,\AA}$) and scattering length density (SLD, $\rho$) (in units of $\times 10^{-6}{\textup{\,\AA}}^{-2}$). As no enrichment was performed, the natural isotopic abundance was assumed in the current work, in which case the SLD can be determined from the standard mass density and molar mass of each component: 
\begin{equation}
    \rho_N =N_A\sum_{i}^{}\frac{p_i}{A_i}<b_i>,
\end{equation}
where $N_A$ is Avogadro's constant, $p_i$ is physical density, $A_i$ is atomic mass, and $<{b_i}>$ is the nuclear scattering length for component $i$ averaged over its natural isotope composition. Each parameter in each layer is varied within physically relevant bounds using a differential evolution algorithm, until the fitted data match the experimental data closely, by minimizing ${\chi}^2$ through a least-squares-curve-fitting method or using Bayesian fitting methods. Prior probability is the probability distribution function for a parameter, $\theta$, given the pre-existing knowledge of a system. The posterior probability is given as $p(\theta|D,I)$, for a system, $I$, and observed data, $D$, and it is consistent with the prior probability and the observed data. It is similar to a least-squares analysis confidence interval. Once a good fit is achieved through the use of the differential evaluation algorithm, a Markov-chain Monte Carlo (MCMC) simulation is performed to examine the posterior probability distribution of parameters, and to assess a family of fits \citep{nelson2019refnx}. The output of the MCMC simulation produces the distribution, spread and covariance of the reflectometry parameters through resampling and refitting from randomized starting points repeatedly as explained in \citep{holt2022using}. For the MCMC analysis 3000 steps (with 500 nburn and 500 equilibration), 200 walkers, and 500 thinning yielded a total of 800 fits. The final parameters presented are the median values from the distribution of all the fits with an uncertainty, that is half of the 15.87 to 84.13 percentile range. 

\subsection{High temperature annealing}
The sample was placed into an enclosed cylindrical high temperature (maximum temperature 1600$^{\circ}\mathrm{C}$) vacuum furnace originally designed at Institut Laue-Langevin (ILL), France. Heat is generated in the furnace by passing a 300 A current through the core of the furnace that has a thin niobium element, the centre of which contains the sample. The furnace walls contain eight thin layers of niobium radiation shields for maintaining a uniform temperature at the sample \citep{goodway2019fast}. The outer aluminum walls of the furnace are highly transparent to neutrons. The neutron transmission through the furnace exceeded 85\%, within the usable wavelength. Base pressures of $10^{-7}$ to $10^{-6}$ mbar were achieved using a turbomolecular pump that pumped the sample volume. A vertical probe stick was used to mount the sample, on a stainless steel (grade 304) holder, which was able to withstand very high temperatures. Variable temperature in-situ neutron reflectometry measurements were conducted with this controlled annealing process, with the temperatures ranging from room temperature to 800$^{\circ}$C. 
\subsection{Thin film characterization XRD}
The crystal structures of both un-annealed and annealed samples were analyzed at room temperature using a Rigaku Smartlab II diffractometer with CuK$\alpha$ radiation, $\lambda$ = 1.540598${\textup{\,\AA}}$. The system employs a $2\theta/\omega$ geometry. Data were collected from $2\theta = 3^{\circ}$ to $80^{\circ}$ with a step of $0.01^\circ$.

\subsection{Magnetization measurements}
DC magnetization measurements were performed using a vibrating sample magnetometer (VSM) inserted in a physical property measurement system (PPMS) from Quantum Design. The temperature dependence of the sample magnetic moment m(\textit{T}) was measured using the zero-field cooling (ZFC) method at different magnetic fields up to 80 kOe. Also, the isothermal magnetization curves, ($\textit{m}-\textit{H}$), were measured for temperatures over a range from 3 K to 15 K, which are well below to slightly above the superconducting transition temperature, $\textit{T}_{\text{c}}$. All the magnetization measurements in this study were performed with the magnetic field aligned perpendicular to the sample’s surface.

After the neutron reflectometry measurements, in-situ annealed samples were diced into small pieces ($\sim$3 mm$^2$) and were used for the magnetization measurements. The sample was mounted on to  the VSM sample holder, which has  a small brass trough with quartz braces for holding the sample.  The sample was sealed in place with Teflon tape after placing it between the quartz braces. The VSM holder is further mounted on to the probe, which was inserted into the VSM for offset measurements to specify the sample position in the VSM, and for subsequent measurements. Here, the ZFC is  achieved by rapidly cooling the sample at the rate of 20 K/min in zero magnetic field from a temperature of 300 K down to 3 K. Constant magnetic fields were applied (100, 500, 1000, and 10000 Oe) perpendicular to the sample’s surface and the temperature was increased at the rate of $\sim$0.9 K/min from 3 K to 30 K, for the m(\textit{T}) measurements. When the temperature reached 30 K for each of the measurements, the field strength was set to zero, then the sample was cooled down to 3 K and the measurement was repeated for the next field strength. Similarly, for each isothermal magnetization measurement, the sample was cooled down to 3K at zero field strength, and  the field was swept from 0 to 80 kOe at the rate of $\sim$1000 Oe/min.

Henceforth in this paper, `Sample A' will refer to the sample in-situ annealed during the neutron beam reflectometry measurements at 750 $^{\circ}$C, while `Sample B' will denote the sample annealed at 800 $^{\circ}$C.

\subsection{Thin film characterization EDX/TEM}

To characterize the samples after the in-situ annealing, transmission electron microscopy (TEM) was employed. To do this, the sample surfaces were first coated with a 30~nm carbon layer. Following this, the area of interest on the surface (20 $\mu\text{m}^2$) was protected by first depositing a thin layer of platinum (Pt) film using an electron-beam, followed by a thicker layer using a focused ion beam (FIB) system. Cross-sectional lamellae ($\sim$100 nm) were then prepared utilizing a 30 keV focused ion beam through a lift-off technique, followed by a final etching with a 5 keV gallium (Ga) FIB to minimize any surface damage. Both cross-sectional TEM and high-angle annular dark-field scanning transmission electron microscopy (HAADF-STEM) images were acquired in scanning electron microscopy (SEM) mode using an FEI Tecnai TF20 TEM operating at 200 kV.  The stoichiometry of the thin film layers in the samples was further analyzed using energy dispersive X-ray spectroscopy (EDX) with an $\sim$1 nm diameter probe beam (EDAX Apollo XLT 2 detector).

\section{Results and Discussion}
\label{secIII}
\subsection{Reflectometry results}
\label{sec:reflectometryresults}
\begin{figure}[!htb]
  \centering
  \includegraphics[trim=0 2 30 20, clip, width=1\linewidth]{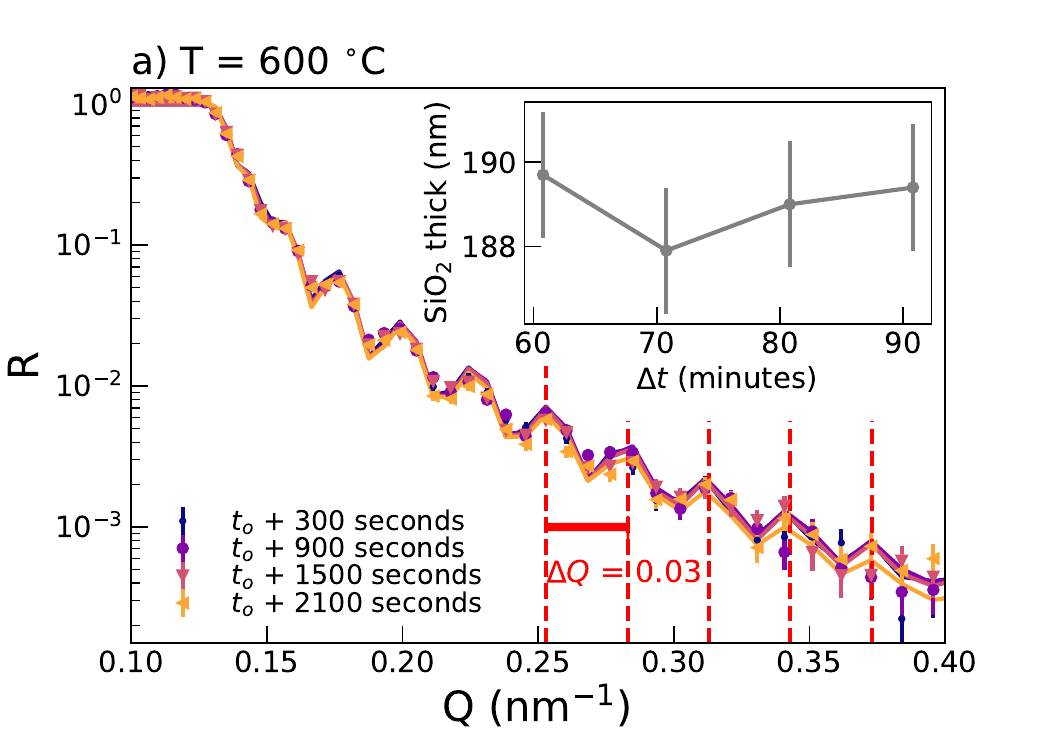} \newline
  \includegraphics[trim=0 2 30 20, clip, width=1\linewidth]{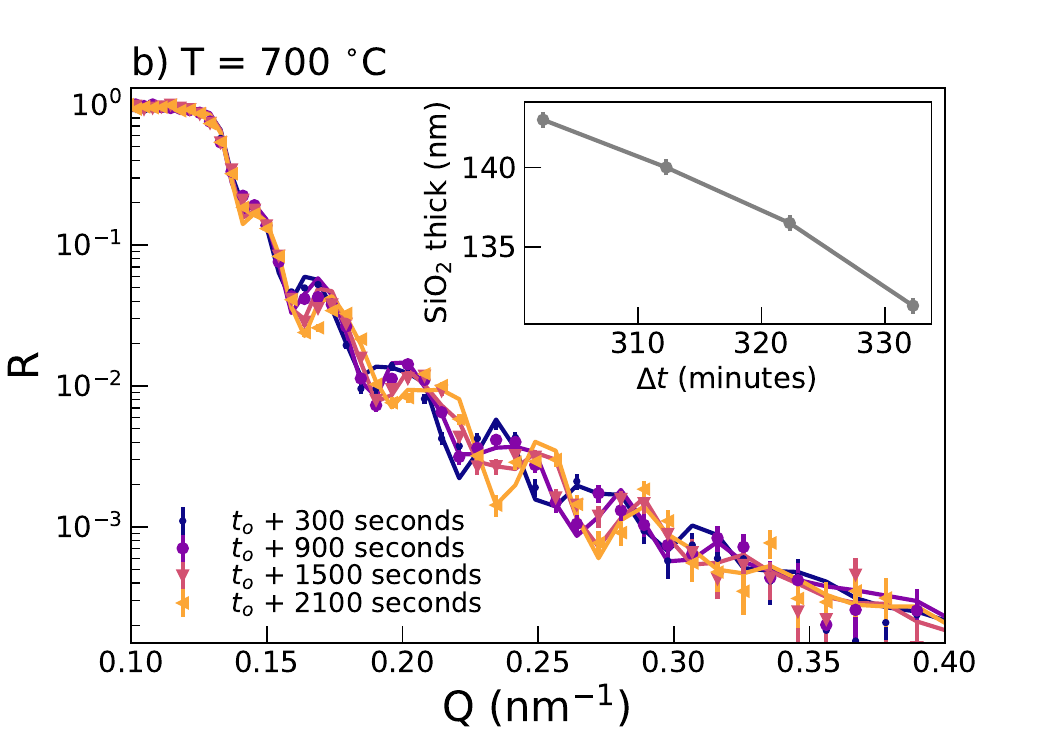}   \newline
  \includegraphics[trim=0 2 30 20, clip, width=1\linewidth]{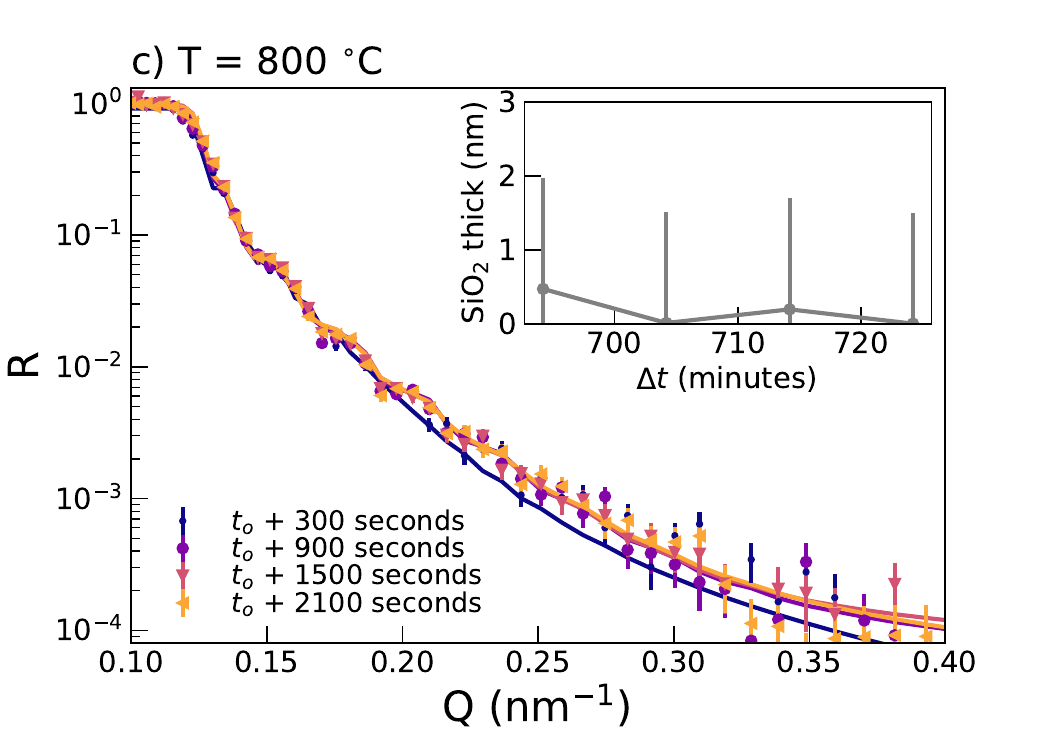}
  
  \caption{Time-dependency of the neutron reflectivity experimental data for Sample B at a) 600 $^{\circ}$C, b) 700 $^{\circ}$C and c) 800 $^{\circ}$C. The points are the experimental data, and the solid lines are the fits. $t_0$ is defined as the time since the target temperature has been reached, alignment has been achieved and the isothermal run has started. The inset shows the silica thickness with $\Delta t$ being the time since the furnace exceeded 600 \degC.}.
  \label{fig:lowQ_Ref_TimeDep}
\end{figure}
\raggedbottom

Time-resolved reflectivity patterns were obtained at multiple temperatures for the two samples each heated to a maximum temperature of 750 $^{\circ}$C and 800 $^{\circ}$C respectively. 

Chemical reactions at the SiO$_2$/V interface consume Si and O, leading to a modified SiO$_2$ layer thickness, and enabling the growth of secondary VO$_x$ and V$_3$Si phases. The time-dependency of the neutron reflectivity patterns at different temperatures can be used to identify the start, end and transition rate of the chemical reaction. The main observables in the patterns are the high frequency Kiessig fringes associated with the thick oxide layers (initially SiO$_2$). As the reflectometry patterns are, to a first approximation, the 1D Fourier transforms of the real-space atomic (nuclear) structure, the Kiessig fringe periodicity can be used to estimate the layer thickness ($d$)  via the relation 
\begin{equation}
\label{eq:BornApproximationThicknessFormula}
    d = \frac{2\pi}{\Delta Q}
\end{equation} 
where $\Delta Q$ is the spacing between two fringe maxima.  Figure \ref{fig:lowQ_Ref_TimeDep} a) shows multiple reflectometry patterns, each measured over ten minutes, at 600 \degC\, where  $\Delta Q$ has been labeled. The fringe spacing at 600 \degC\ is $\Delta Q \approx 0.03$ nm$^{-1}$ which is consistent with a SiO$_2$ layer of 200 nm. It is well known that the latter approximation is not exact, due to dynamic scattering effects which are stronger at low Q. Using the full-model (described later, which includes dynamic scattering) the determined starting thickness is 189 $\pm$ 5 nm. In the figure, the points are the experimental data and the solid lines are fits to the data using the SLD model described in the following section. The variation of the fringe spacing ($\Delta Q$) provides a good first indication of the variation of the oxide layers during the growth. For example, Figure \ref{fig:lowQ_Ref_TimeDep} b and Figure \ref{fig:lowQ_Ref_TimeDep} c show the time dependency for 700 $^{\circ}$C and 800 $^{\circ}$C over a series of ten-minute intervals. The data for 700 $^{\circ}$C show a clear time dependency, indicating interfacial chemical reactions which lead to a modified SiO$_2$ thickness that shifts the high frequency fringes.

In contrast the patterns at 600 $^{\circ}$C and 800 $^{\circ}$C are independent of time. The inset shows the fitted SiO$_2$ thickness. At 600 \degC, the temperature was too low for the reaction to proceed on observable timescales so the SiO$_2$ thickness is effectively constant (189 $\pm$ 5 nm) over the first hour. Between 700 and 800 \degC, the SiO$_2$ thickness dropped dramatically. By 800 $^{\circ}$C, the SiO$_2$ layer was fully consumed at the start of the measurement time, and the main reaction stopped. Whilst some high frequency fringes are present in the 800 \degC\ data, these are heavily damped in comparison with those at 600 \degC, and also appear at different Q-points. Detailed modelling in the next section shows these are not from the SiO$_2$ but instead mainly arise from the rough, vanadium oxide surface layer that is formed by the reaction.

By modelling the full intensity of the reflection patterns as a function of Q including dynamic scattering effects, the scattering model also enables measurement of structural parameters, including interface roughness (diffusion profiles). The model is also able to discriminate between the thickness of the SiO$_2$ and VO$_x$ layers which is not possible by simply applying Equation \ref{eq:BornApproximationThicknessFormula}. The fitting procedure provides quantitative values for the layer thicknesses during the growth, and therefore the reaction kinetics, at different times and temperatures.  In order to fit the evolution of the thin film reflectivity data, a self-consistent chemical model was developed that imposed constraints on the scattering length density (SLD) profile. Constrained fitting was necessary because direct Fourier inversion was not possible, as is generally the case in neutron scattering, since the observed intensity signal is the modulus of the neutron wave function, $|\psi|^2$, and phase information is lost. A model is therefore needed to fit the data in this system. 

Modeling is complicated, however, by the very low  coherent neutron scattering cross-section of vanadium metal film there is a low contrast compared to  those of the oxide layers. Additionally, while it was possible to measure to high Q for the time-independent datasets below 600 \degC\ and above 800 \degC,  only low Q data could be reliably fitted at 700--750 \degC\, due to the sample kinetics which necessitated short integration times, such that only the high intensity regions of the reflectivity curve could be observed with acceptable statistical uncertainty.  To yield reliable fits using a self-consistent chemical model, the initial layer thicknesses were determined prior to the reaction, and a mass-balanced model was implemented to fit the oxide and V$_3$Si layer thicknesses during the reaction. The model assumes that the key chemical reactions are of the general form:
\begin{equation}
xSiO_2 + yV \rightarrow xV_3Si + (y-3x)VO_{2x}
\end{equation}
where $y > x$. 
Based on trial fitting, and consistent with the XRD results, we found that the specific reaction ($x$=3, $y$= 13) describes the data very well at most temperatures, via the reaction:
\begin{equation}
    3SiO_2 + 13V \rightarrow 3V_3Si + 2V_2O_3
\end{equation}
The method assumes that the layer volume, determined by film area and thickness ($d_{SiO2}$) of the SiO$_2$ is related to the amount of SiO$_2$ used in the reaction. The number of SiO$_2$ units used in the reaction per unit area is:
\begin{equation}
  \Delta x  = (d_{SiO_2}-d_{SiO_2}') \times N_{Si}^{SiO_2}   
\end{equation}
where $d_{SiO2}=d_{SiO2}(t,T)$ is the thickness of the layer at a temperature ($T$) and time ($t$); $d'$ is the thickness at room temperature at t=0, before the reaction. $N_{Si}^{SiO2}$ is the number density of Si atoms per volume  which is determined from the mass density $\rho_{SiO_2}$ and molar mass $M$ such that $N_{Si}^{SiO2} = \rho_{SiO2}/(N_A M_{SiO2})$. Once $\Delta x$ is determined, this then fixes the amount of V$_3$Si and V$_2$O$_3$ produced via the chemical reaction, and this therefore determines the thicknesses of the V$_3$Si and V$_2$O$_3$ layers via the relations:
\begin{equation}
  d_{V_3Si} = \Delta x / N_{Si}^{V3Si}
\end{equation}
\begin{equation}
  d_{V_2O_3} = 2 \times \Delta x / N_{O}^{V_2O_3} 
\end{equation}
where $N_{Si}^{V3Si}$ is the number density of $Si$ atoms in the V$_3$Si phase, and $N_{O}^{V2O3}$ is the number density of oxygen atoms in the V$_2$O$_3$ phase. Finally, to balance the chemical reaction, the  thickness of any left-over (unreacted) vanadium is:
\begin{equation}
d_{V} = d'_{V}  - (3\Delta x  + \frac{4}{3}\Delta x) \times N_{V}^{V} =   d'_{V} - N_{V}^{V} \frac{13}{3}\Delta x 
\end{equation}
where $d'_{V}$ is the starting V thickness prior to the reaction; $N_{V}^{V}$ is the number density of vanadium atoms in the metallic vanadium phase and the (3)   appears because the ratio V:Si in V$_3$Si is 3:1, and the factor 4/3 appears because the ratio V:O in V$_2$O$_3$ is 2:3, multiplied by 2 as each reacted SiO$_2$ formula unit releases 2 oxygen. 

Using this chemical model to guide the fitting procedure, the four thickness parameters in the reflectivity model ($d_{SiO_2}$, $d_V$, $d_{V_3Si}$ and $d_{V_2O_3}$) are effectively constrained so that they all depend on a single free parameter ($\Delta x$) which greatly improves the reliability of the fitting.  One additional variable was introduced to model a thin vanadium native oxide layer caused from ambient exposure prior to heating. However this was constrained to a thickness of $< 2$ nm and had a negligible effect on the fits overall. 

\begin{figure}[htb]
  \centering
  \includegraphics[trim=1 10 35 40, clip, width=1\linewidth]{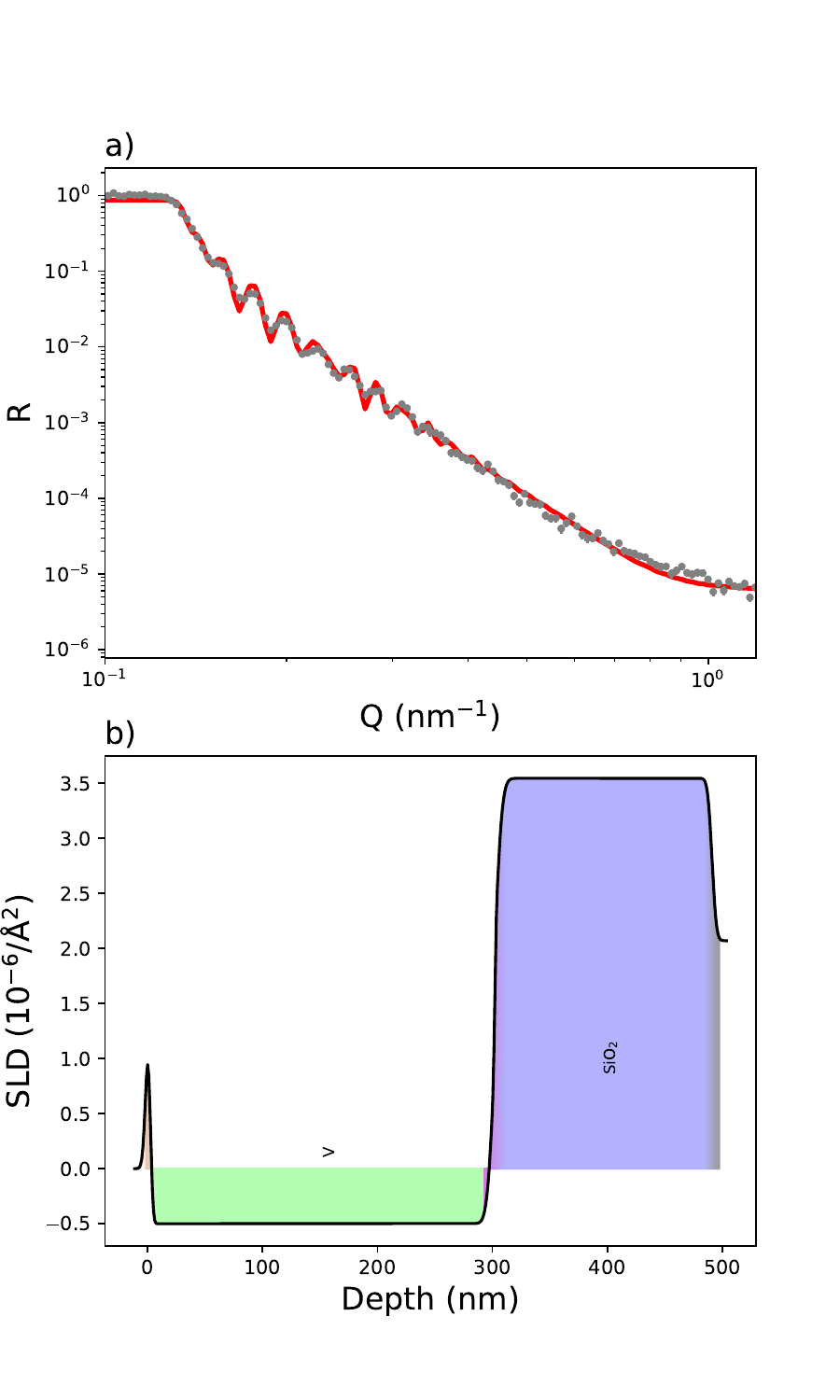}
  \vspace*{-10mm}
  \caption{a) Logarithmic-linear representation of the neutron reflectivity experimental data as a function of Q, obtained during the in-situ annealing of V samples at a temperature of 600 $^{\circ}$C. In the plot, the solid circles represent the experimental data, and the red solid lines fits to the data.  b) The best-fit SLD profile.  \label{fig:Ref600}}
\end{figure}

To test the accuracy of the above model before, after and during the reactions, the data were fitted at multiple temperatures. Initially the time-independent dataset at 100 \degC\ was fitted to determine the virgin state of the sample, with $d_{SiO_2}$ = 195 $\pm$ 5 nm at $\Delta x = 0$.  Figure \ref{fig:Ref600} shows the fitted reflectivity patterns and SLD profile at 600 \degC. This fit shows that the V and SiO$_2$ films are both present as individual layers at 600 \degC, similar to at room-temperature, with minimal reaction having occurred. 

\begin{figure}[!htb]
  \centering
  \includegraphics[trim=0 10 35 40, clip, width=1\linewidth]{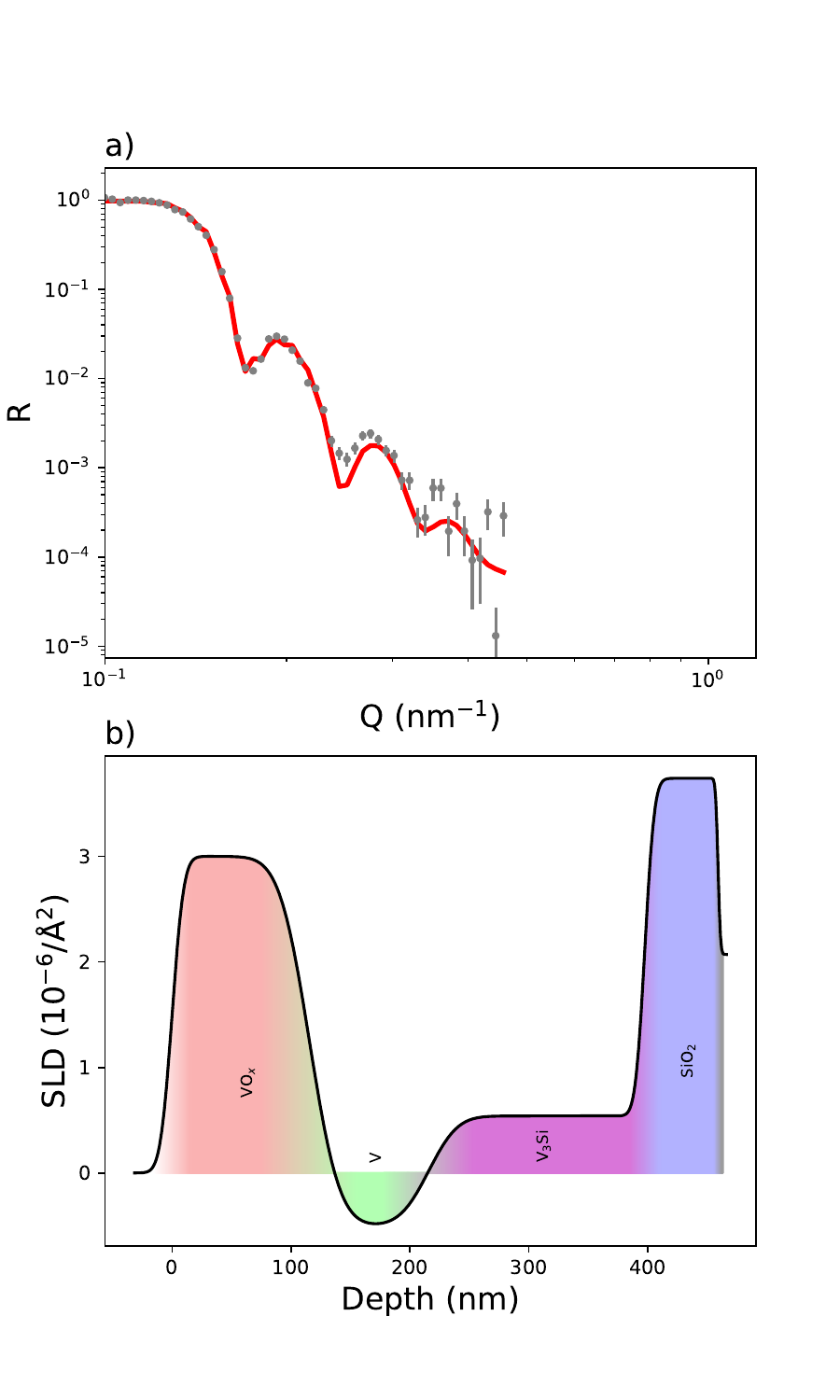}
  \vspace*{-10mm}
  \caption{a) Logarithmic-linear representation of the neutron reflectivity experimental data as a function of Q, obtained during the in-situ annealing of V samples at a temperature of 750 $^{\circ}$C. In the plot, the solid circles represent the experimental data, and the red solid lines fits to the data.  b) The best-fit SLD profile.}
  \label{fig:Ref750}
\end{figure}

\begin{figure}[!htb]
  \centering
  \includegraphics[trim=0 10 35 40, clip, width=1\linewidth]{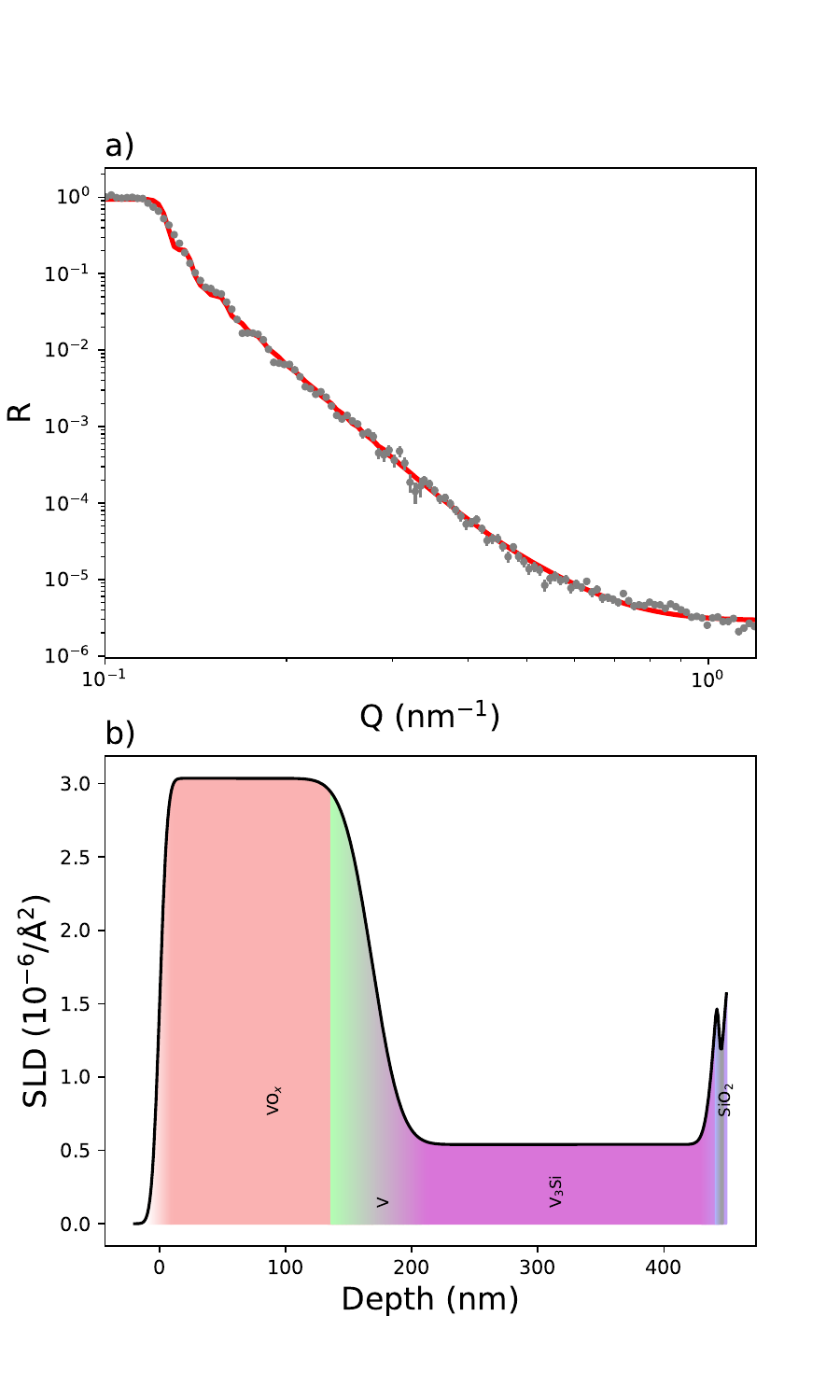}
  \vspace*{-10mm}
  \caption{a) Logarithmic-linear representation of the neutron reflectivity experimental data as a function of Q, obtained during the in-situ annealing of V samples at a temperature of 800 $^{\circ}$C. In the plot, the solid circles represent the experimental data, and the red solid lines fits to the data.  b) The best-fit SLD profile.}
  \label{fig:Ref800}
\end{figure}

In contrast, Figure \ref{fig:Ref750} shows the reflectivity patterns and SLD profile at 750 \degC\, indicating a modified chemical profile after several hours of reaction. Figure \ref{fig:Ref800} shows the fitted reflectivity patterns and SLD profile at 800 \degC\ after the reaction had run to conclusion over several hours. In all cases the model produces an excellent description of the data. The SLD profiles also provide insights into the diffusion and roughness parameters at each temperature. For comparison, the theoretical SLDs for the single phases, assuming the bulk densities, are reported in Table \ref{table:SLD}. An interesting situation was observed near the midpoint of the reaction at 750 \degC, as shown in Figure \ref{fig:Ref750} where the thickness of the SiO$_2$ and V$_2$O$_3$ both became comparable and relatively thin films (60 nm) formed. This is fully described by the model above. After the reaction has completed at 800 \degC, the SiO$_2$ layer was fully consumed, and best-fit to Sample B yielded a thick V$_3$Si layer (265 $\pm$ 5 nm) and thick oxide layer (170 $\pm$ 5 nm,), consistent with the post-heat treatment TEM data presented later.
\vspace{-1mm}

\raggedbottom

\begin{center}
\begin{table}
\caption {Theoretical scattering length densities \label{table:SLD}}
\begin{tabular}{|c | c | c | c|} 
 \hline
 Compound & Mass density & X-ray SLD & Neutron SLD\\ 
   &  (g/cm$^3$) & (10$^{-6}$/ \r{A} $^2$) &(10$^{-6}$/ \r{A} $^2$)\\ [0.8ex] 
 \hline
 Si & 2.33 & 20.07 & 2.07 \\ 
 \hline
 SiO$_2$ & 2.2 & 18.86 & 3.47 \\
 \hline
  V & 6.11 & 46.97 & -0.32 \\
 \hline
 V$_2$O$_5$ & 3.35 & 27.01 & 3.12 \\
 \hline
 V$_2$O$_3$ & 4.87 & 38.77 & 3.23 \\ 
 \hline
 VO$_2$ & 4.65 & 37.27 & 3.77 \\ 
 \hline
  V$_3$Si & 5.77 & 36.12 & 0.54 \\
 \hline
 V$_5$Si$_3$& 5.29 & 41.89 & 0.96 \\ 
 \hline
 VSi$_2$ & 4.42 & 36.12 & 1.95 \\
 \hline
\end{tabular}
\end{table}
\end{center}
\vspace*{-10mm}

The approximate growth rates observed during the experiment, matching the time regions listed in \mbox{Figure \ref{fig:lowQ_Ref_TimeDep}} are reported in Table \ref{table_rates}. Here we have used a linear fit to the thicknesses observed over these measured time windows. The table indicates the rate of the depletion (negative thickness ``gain") of the SiO$_2$, and the associated positive thickness-gain of the other layers. It is worth stating that the growth rate of a diffusion-mediated reaction is generally non-linear over long time spans, and the rate depends on the amount of SiO$_2$ which evolves over time. Thus,  the latter rates are simply an indication of the growth observed at particular points during the experiment, provided as a guide for future work.

In general, while the chemical model describes the relative amounts of the different phases, there is still the question of whether those phases form discrete layers and the spatial position of each layer. Reflectivity addresses this question as trial-fitting showed only certain SLD profiles fit the data well. In general, for $T > 750\,^{\circ}\mathrm{C}$ the reflectivity profiles only could be well-described with the profile where the vanadium oxide formed on top of the silicide as a discrete layer. This also concurs with the TEM data. The interpretation is that the different surface energy/surface tension of the two phases favours the latter configuration. At an earlier stage of the growth (650 \degC\ and early times at 700 \degC) there is some evidence, however, for mixed layer formation of a  silicide/vanadium oxide above the SiO$_2$ interface, which warrants investigation in future work.
\vspace{-10mm}

\begin{center}
\begin{table} 
\caption {Observed film growth rates at different temperatures \label{table_rates}}
\begin{tabular}{|c | c | c | c|} 
 \hline
 Temperature & SiO$_2$ & V$_3$Si  & V$_2$O$_3$\\
(\degC)& (nm/min)& (nm/min) & (nm/min)\\ [0.8ex] 
 \hline
 600 & 0 & 0 & 0 \\ 
 700 & -0.11 & 0.15 & 0.10 \\ 
 750 & -0.39 & 0.55& 0.34 \\ 
 \hline
\end{tabular}
\end{table}
\end{center}
\raggedbottom

\subsection{XRD Analysis}

\begin{figure}[!htb]
  \centering
  \includegraphics[trim=30 25 30 25, clip, width=1\linewidth]{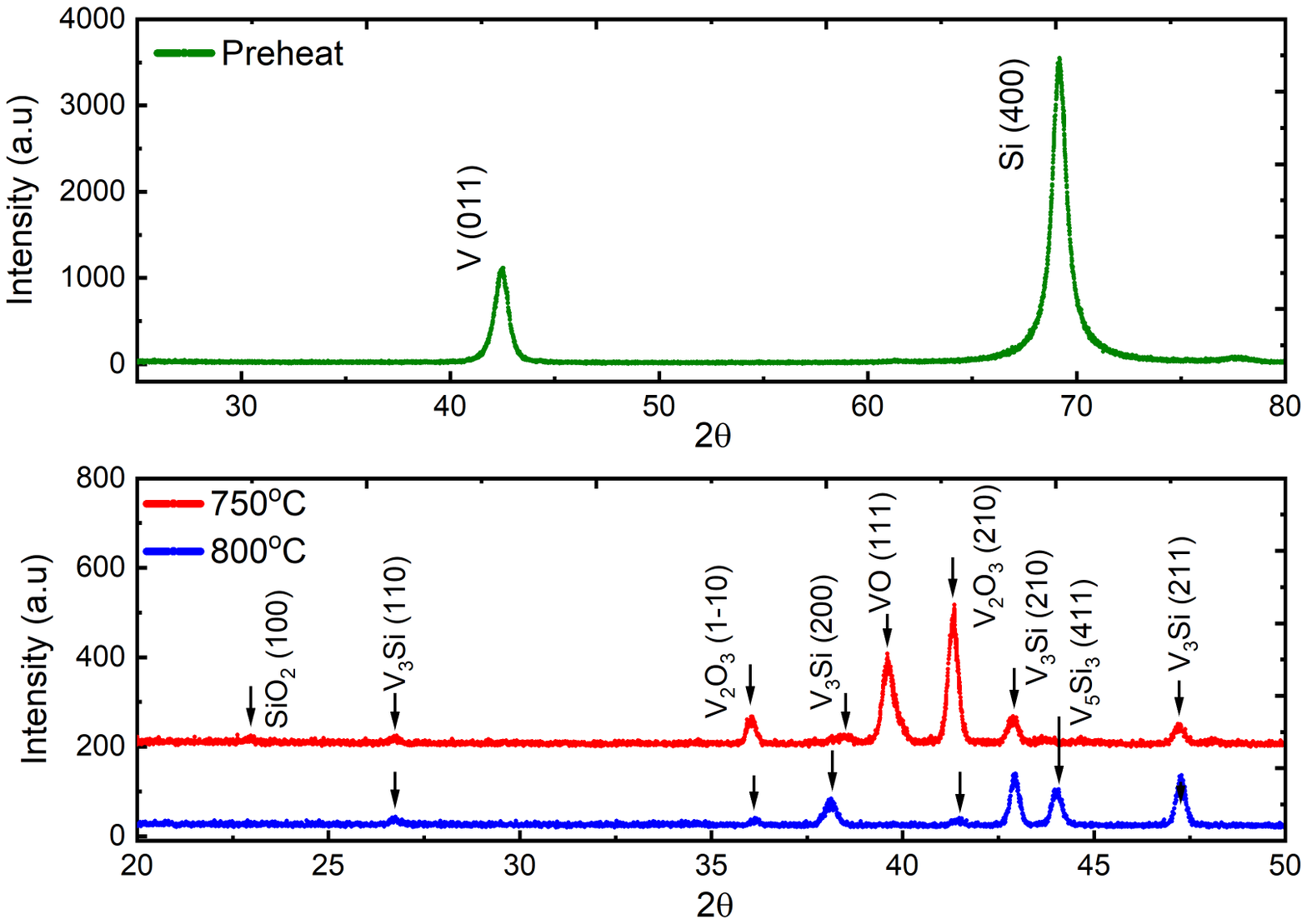}
  \caption{X-ray diffraction patterns of an un-annealed sample (Preheat), sample A (750 $^\circ$C) and sample B (800 $^\circ$C)}
  \label{xrd}
\end{figure}

X-ray diffraction (XRD) patterns for three samples, an un-annealed sample (Preheat), Sample A annealed at 750 $^{\circ}$C, and Sample B annealed at 800 $^{\circ}$C, are presented in Figure \ref{xrd}. The un-annealed sample displays diffraction peaks that are characteristic of metallic vanadium deposited on a silicon substrate that has a thermally oxidized SiO$_2$ layer. The diffraction pattern for Sample A, which underwent in-situ annealing during the neutron reflectometry study, confirms the formation of V$_3$Si by exhibiting all major peaks associated with this phase. Also, peaks indicative of V$_2$O$_3$ and VO are observed. In contrast, the XRD pattern for Sample B, also prepared similarly, but annealed at 800 $^{\circ}$C shows a predominance of V$_3$Si peaks with a noticeable reduction in the intensity of the oxide peaks compared to the oxide peaks for Sample A. It is also observed that the SiO$_2$ peaks have disappeared, and an additional V$_5$Si$_3$ phase has formed, indicating that the SiO$_2$ bottom layer was completely consumed during the reaction, exposing the underlying silicon substrate for further reaction. The reaction of remaining metallic vanadium with the silicon substrate is suggested to be the cause of the formation of this additional phase. 

\vspace{-5mm}
\subsection{TEM Analysis}
\label{TEM}
\begin{figure}[!htb]
  \centering
  \includegraphics[width=1\linewidth]{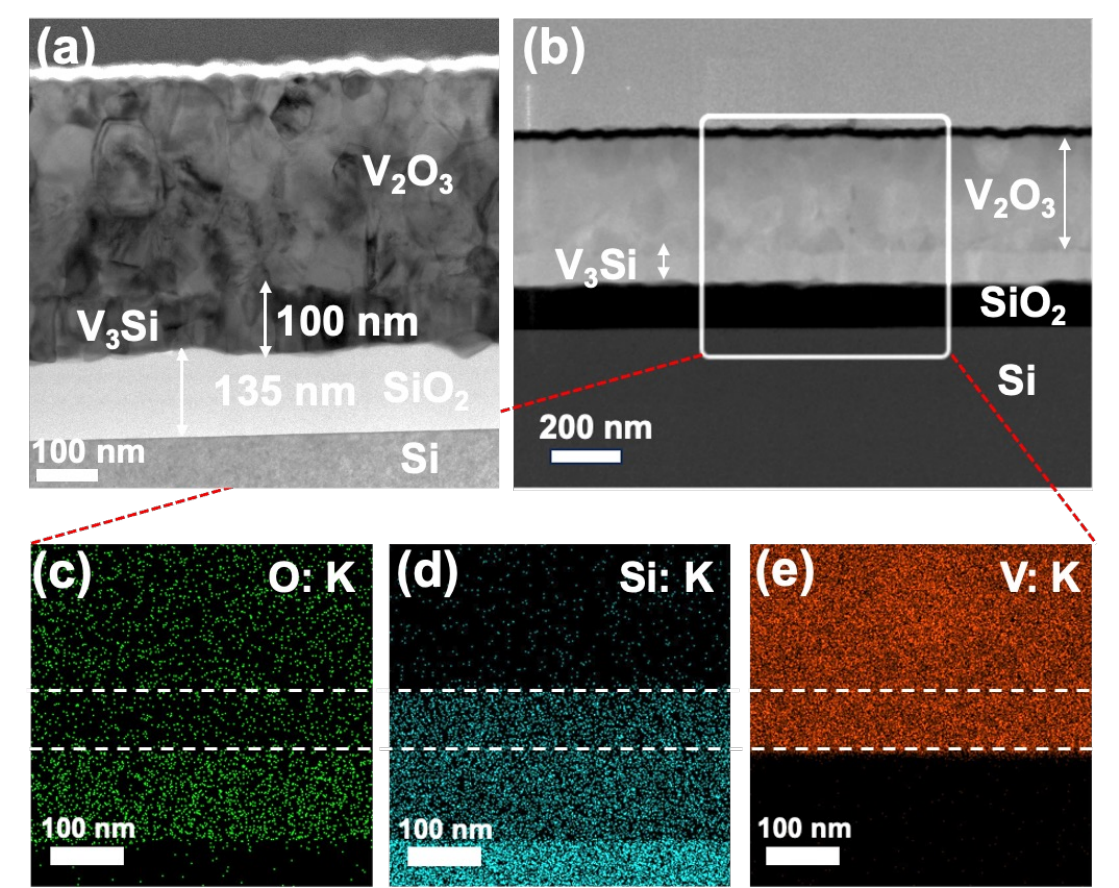}
  \vskip 0.3cm 
  \includegraphics[width=1\linewidth]{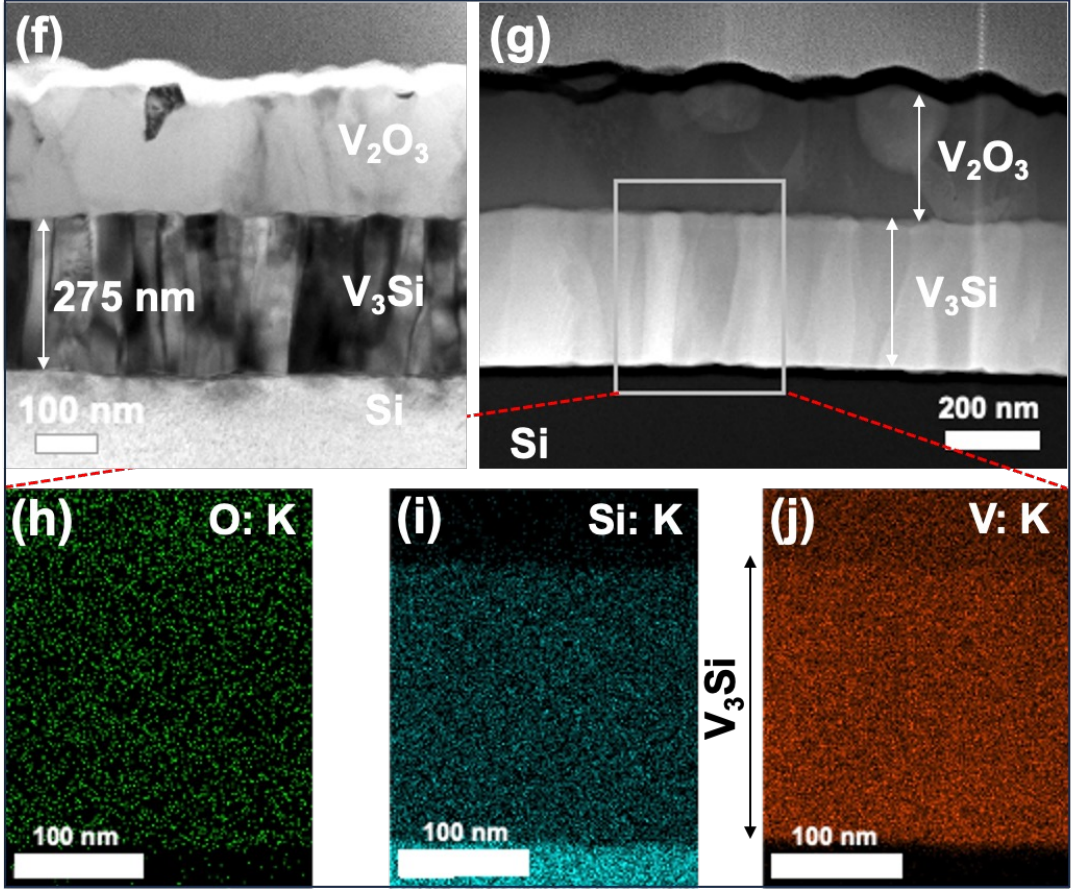}
  \caption{Cross-sectional TEM images of Sample A (a) and Sample B (f) reveal the V$_3$Si silicide layer, its interfaces, the columnar grain structures, and the oxide layer atop the V$_3$Si. HAADF-STEM images acquired in SEM for Sample A (b) and Sample B (g) further illustrate the structural composition. EDAX elemental mapping for oxygen, silicon, and vanadium, derived from the HAADF images, as shown for Sample A (c-e) and Sample B (h-j), highlight the elemental distribution across the samples.}
  \label{figTEM}
\end{figure}

Figure \ref{figTEM} shows the cross-sectional TEM images, HAADF-STEM images and EDAX mapping  of sample A (a-e) and sample B (f-j). Both samples exhibit a continuous V$_3$Si layer with an oxide layer on top, formed due to the diffusion-controlled growth process. The thicknesses of the V$_3$Si layers are approximately 90 nm for sample A, and 280 nm for sample B. The vanadium oxide layers on top are 320 nm thick for sample A, and 180 nm for sample B. It is to be noted that the bottom SiO$_2$ layer in sample B is fully consumed during the annealing process, leaving the V$_3$Si layer directly on the Si substrate. This result is consistent with the reflectometry model presented in \ref{secIII}.\ref{sec:reflectometryresults} for the sample B, which further validates its accuracy. Careful analysis of the V$_3$Si morphology reveals a columnar grain structure in both the samples, characterized by elongated vertical grains growing perpendicular to the plane of the film surface.

The role of the columnar grains can be understood from the flux pinning mechanism at the grain boundaries. Columnar grain growths are reported in A15 superconductors like V$_3$Ga \citep{livingston1977} and Nb$_3$Sn \citep{godeke2005}, where the superconducting phase formation is through a diffusion process. The average widths of grains are estimated to be 35 nm for sample A, and 60 nm for sample B in this study. It is clearly seen that the grain size increases with the annealing temperature from 750 $^{\circ}$C to 800 $^{\circ}$C, and the columnar grains become more prominent with the increase in temperature.

\subsection{Magnetization results}

\begin{figure}[!htb]
  \centering
  \includegraphics[width=1\linewidth]{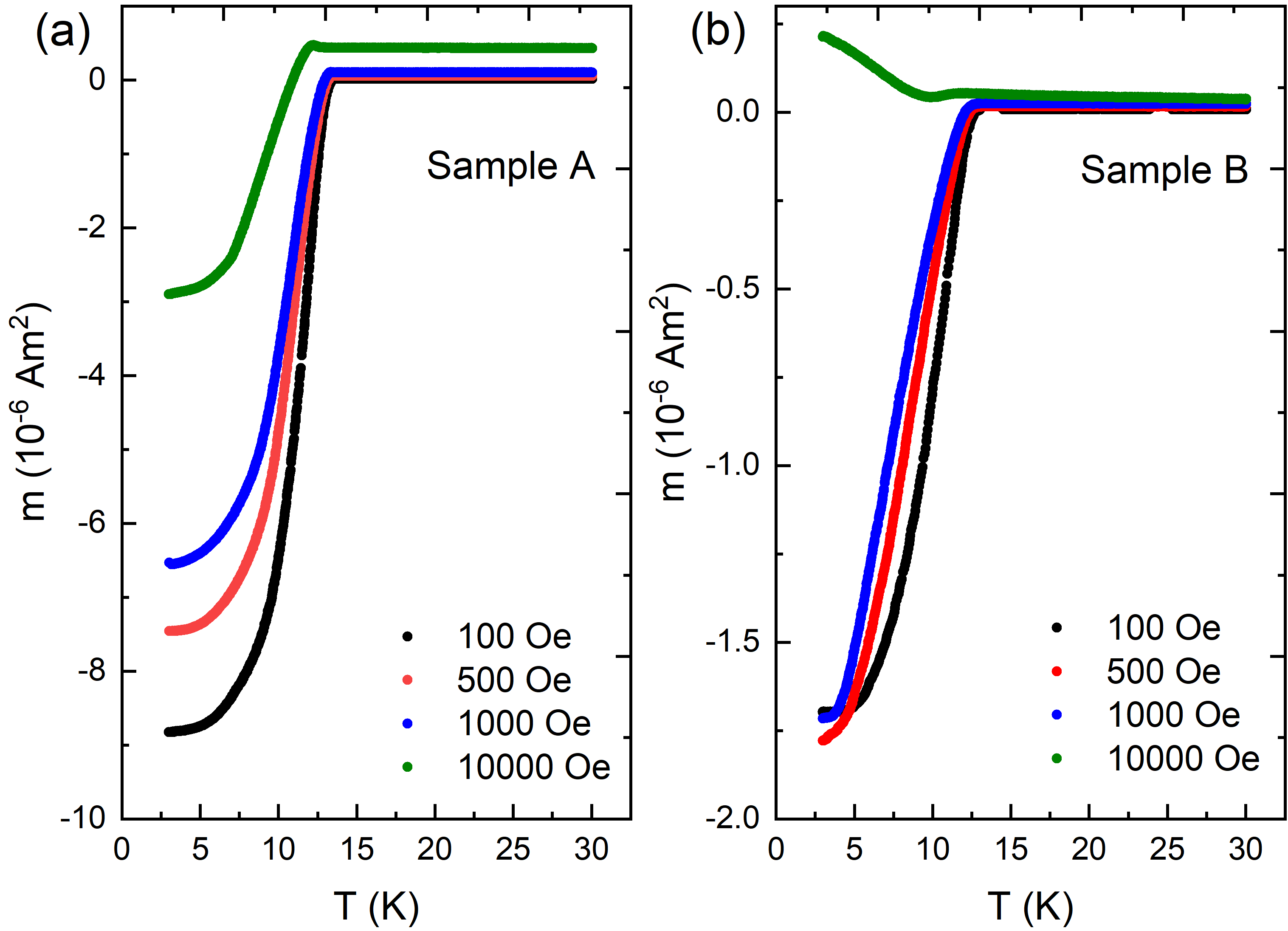}
  \caption{Temperature-dependent magnetization measured at various magnetic fieeds for (a) sample A and (b) sample B.}
  \label{fig1}
\end{figure}

Figure \ref{fig1} shows the dependence of the DC magnetic moment on temperature, for the in-situ annealed samples A and B, respectively. Each sample was ZFC to \mbox{3 K}, and then warmed under constant magnetic fields of 100, 500, 1000, and 10,000 Oe, during which the magnetic moment ($\textit{m}$) was measured. For both the samples, the onset of superconductivity, $\textit{T}_{\text{c}}$(onset) was determined to be \mbox{$\sim$13 K}. In this study, the onset temperature is considered to be the critical temperature. There are studies that considered the midpoint of the diamagnetic transition as $\textit{T}_{\text{c}}$; but this approach has a few challenges \cite{Goldfarb1991}. It should be noted that while bulk single crystal V$_3$Si has a $\textit{T}_{\text{c}}$ of 17.1 K, that of V$_3$Si thin films varies based on growth conditions such as the substrate material \citep{Hauser1963,Howard2023}, deposition method \citep{Bangert1985,ETCroke1988}, and annealing temperature \citep{Nava1_1986}. V$_3$Si films are commonly developed on substrates like silicon, quartz, graphite, magnesium oxide, sapphire, copper, and niobium. In reported results, the highest $\textit{T}_{\text{c}}$ in V$_3$Si thin films ($<$1 $\upmu$m thick) ranges from 10 to 16 K, depending on the growth conditions. Such a decrease in the $\textit{T}_{\text{c}}$ of thin films compared to the bulk, could be explained by the proximity effect induced by the substrate on the V$_3$Si films \citep{clarke1968}. 

The magnetization in the superconducting state is saturated for both the samples at the lowest measured temperature. Below $\textit{T}_{\text{c}}$, the diamagnetism is enhanced due to screening supercurrents. There is a decrease in flux expulsion as external magnetic field strength increases, which is a characteristic of the mixed state for type-II superconductors \citep{abrikosov2004}, and this results in the suppression of a superconducting response.  Hence, there is a significant reduction in superconducting volume that is observed with the increase in magnetic field. Notably, in sample B at 10 kOe, a paramagnetic Meissner effect or Wohlleben effect was observed \citep{geim1998,kostic1996}. In both samples A and B, diamagnetic saturation is observed only below 5 K. This broad transition in the magnetization curves could be a result of strong vortex pinning, thermal fluctuations leading to flux creep and inhomogeneous pinning distributions in the film \citep{vinokur1990}. This indicates that there is irreversibility in the magnetization. 

\begin{figure}[!htb]
  \centering
  \includegraphics[width=1\linewidth]{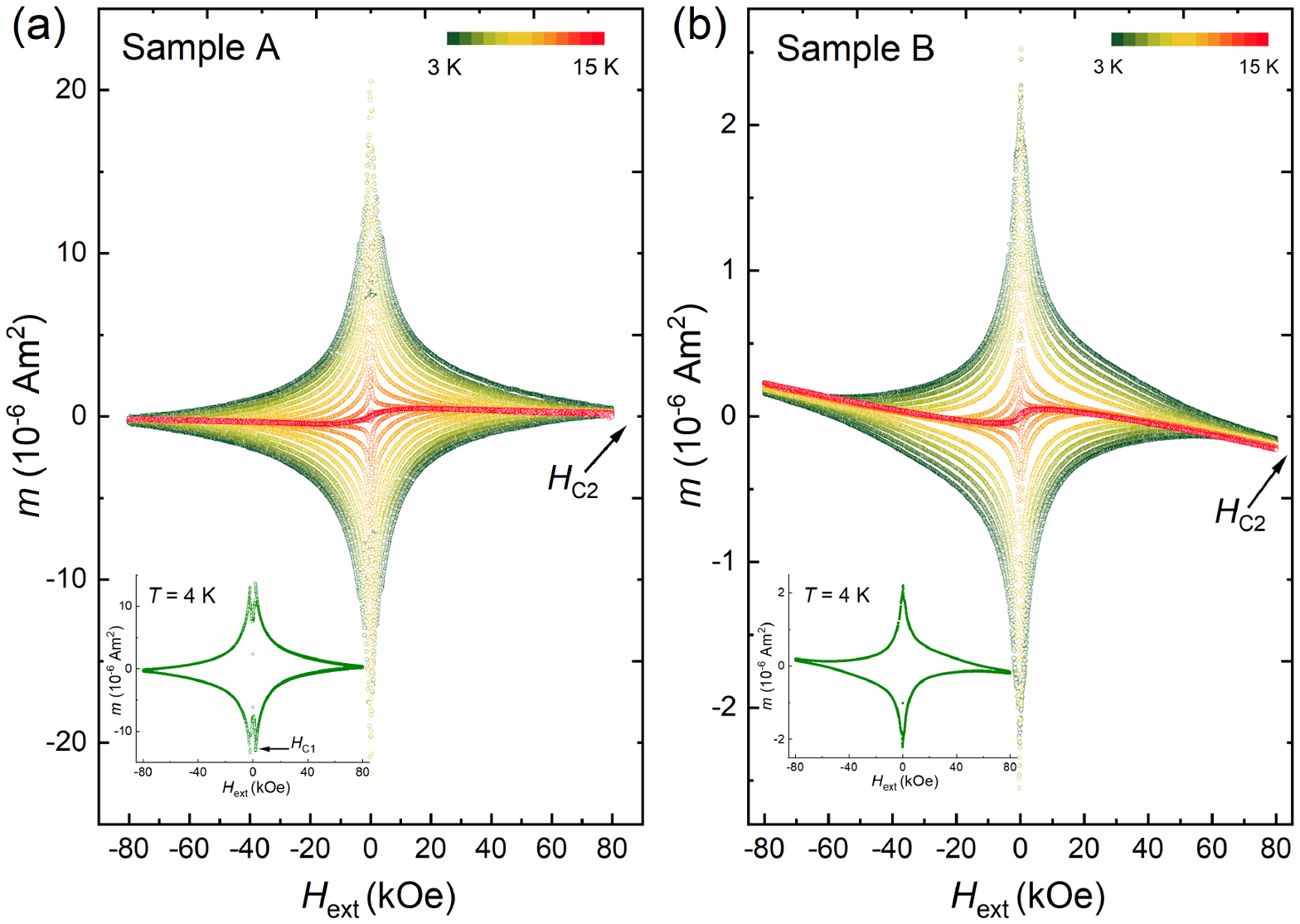}
  \caption{The DC magnetic moment curves as a function of the applied external field $\textit{H}_{\text{ext}}$ in the temperature range of 3 – 15 K for both (a) sample A and (b) sample B are depicted. The insets in both figures display the virgin magnetic moment curves at 3 K. For sample A, the lower critical field ($\textit{B}_{\text{c1}}$) is distinctly defined at temperatures significantly below the transition temperature ($\textit{T}_{\text{c}}$). Both samples demonstrate irreversible magnetization hysteresis behavior below $\textit{T}_{\text{c}}$. A magnetization hysteresis is observed up to 4 T even at 10 K, nearing $\textit{T}_{\text{c}}$.}
  \label{fig2}
\end{figure}

Figure \ref{fig2} illustrates the isothermal DC magnetic moment ($\textit{m}$) as a function of the magnetic field $\textit{H}_{\text{ext}}$ (measured in Oersteds (kOe), for samples A and B at temperatures ranging from 3 K to 15 K, in increments of 1 K. A complete magnetic field cycle, starting from 0 to 80 kOe, returning to 0, then dropping to -80 kOe, and finally reverting to 0, is recorded in the data. It is seen from the isothermal magnetic field dependence of the magnetic moment in both samples A and B that below $\textit{T}_{\text{c}}$, irreversible magnetization hysteresis is observed. As the temperature increases, the width of the magnetic hysteresis loop decreases, resulting in the reduction of the loop area up to $\textit{T}_{\text{c}}$. Since the screening currents are unable to fully expel the trapped flux lines  while the magnetic field is decreased from 80 kOe to 0 Oe, the irreversible magnetic moment results. Even when the external field is reduced to zero, this trapping causes a residual magnetic moment. 

The findings in this study are presented in terms of $\textit{m}$, due to the challenges in accurately determining the volumes of the superconducting material in the samples. But if the volume of the superconducting phase of the films were known, the results could also be expressed in terms of magnetization ($\textit{M}$ = $\textit{m}$/V). The observed hysteresis in the magnetization confirms that the V$_3$Si film formed by this preparation route is a type II superconductor \citep{chaddah1991}. The diamagnetic response of the V$_3$Si layer becomes evident at the start of each trace below $\textit{T}_{\text{c}}$, as the field increases from 0 to 80 kOe, which is consistent with the $\textit{m}$(T) measurement results. The magnetic moment in this region can be expressed as $-\textit{m}(\textit{H}_{\text{ext}})$. 

From Figure \ref{fig1} and \ref{fig2} it can be understood that \mbox{below} $\textit{T}_{\text{c}}$, the magnitude of $\textit{m}$ is larger for sample A than that of sample B. This shows that sample A possess a higher superconducting volume fraction compared sample B, though the V$_3$Si thickness in sample B is larger compared to sample A. Also, in Figure \ref{fig2} (b) for sample B, the magnetic moment curve starts to show a downward bend in high fields close to the critical field, with the increase in temperature. This downward bending of the magnetic moment is possibly due to a weaker superconducting volume fraction formed due to annealing at 800 \degC. This allows the diamagnetic response from the underlying silicon substrate to become more pronounced due to the depletion of SiO$_2$ layer. The thicknesses of the V$_3$Si layers and the depletion of the SiO$_2$ layer in sample B are confirmed through the reflectometry measurements and the cross-sectional TEM analysis discussed in sections \ref{secIII}.\ref{sec:reflectometryresults} and \ref{secIII}.\ref{TEM} respectively.

\begin{figure}[!htb]
  \centering
  \includegraphics[width=1\linewidth]{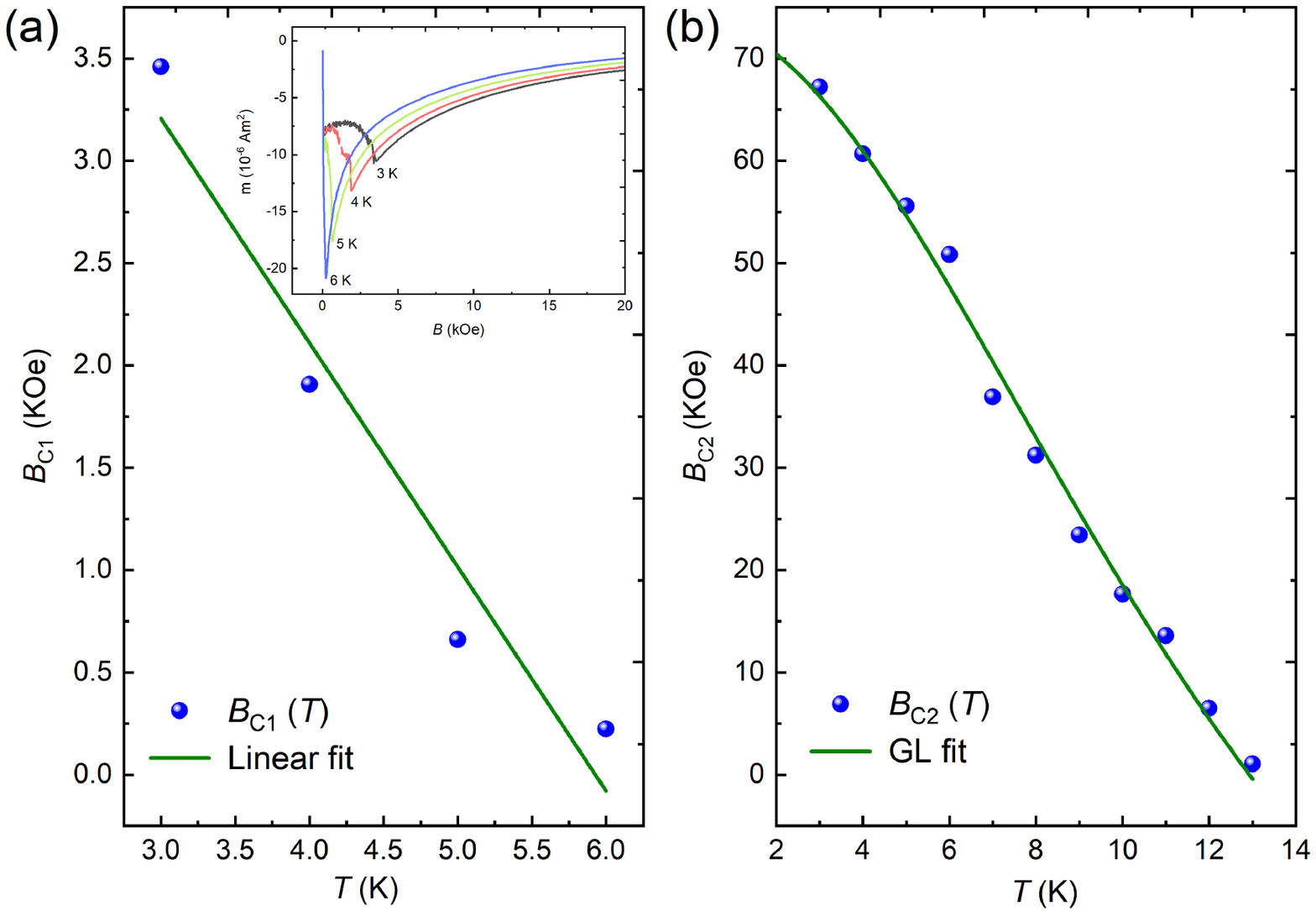}
  \caption{(a) Temperature dependence of lower critical field,  $\textit{B}_{\text{c1}}$, and (b) upper critical field, $\textit{B}_{\text{c2}}$, for the sample A}
  \label{fig3}
\end{figure}

Figure \ref{fig3} shows the temperature dependence of both critical fields, $\textit{B}_{\text{c1}}$ and $\textit{B}_{\text{c2}}$, for sample A. From the plot of magnetic moment $\textit{m}$ versus $\textit{H}$ shown in Figure \ref{fig2}(a), the lower critical field, $\textit{B}_{\text{c1}}$(0) = $\mu_{\text{0}}\textit{H}_{\text{c1}}$, and the upper critical field, $\textit{B}_{\text{c2}}$(0) = $\mu_{\text{0}}\textit{H}_{\text{c2}}$, are derived. For type-II superconductors, $\textit{B}_{\text{c1}}$ is commonly defined as a field at which the initial diamagnetic response on the $\textit{m}$$\textup{--}$$\textit{H}$ curve starts to deviate perfectly from the linear behaviour. As the temperature approaches $\textit{T}_{\text{c}}$, $\textit{B}_{\text{c1}}$ becomes exceedingly small, making its detection near the critical temperature almost impractical due to very subtle deviation, especially in superconductors with high critical current densities \citep{Goldfarb1991}. Figure \ref{fig3}(a) shows the temperature dependence of $\textit{B}_{\text{c1}}$ (in blue solid circle symbol) with respect to temperature, and the inset depicts the $\textit{m}$$\textup{--}$$\textit{H}$ curves measured at low temperatures where the deviation from the linear behaviour of the lower critical field is clearly observable. The olive solid line represents a linear fit to the $\textit{B}_{\text{c1}}$ data points at low temperatures, and the slope -$\textit{dH}_{\text{c1}}$/$\textit{dT}$ is calculated to be $-0.11$ T/K, from which $\textit{B}_{\text{c1}}$(0) $\approx$0.65 T, can be estimated.

The temperature dependence of $\textit{B}_{\text{c2}}$ (in blue solid circle symbol) for sample A is shown in Figure \ref{fig3}(b). $\textit{B}_{\text{c2}}$ at each temperature is estimated from the $\textit{m}$$\textup{--}$$\textit{H}$ plot where the transition from diamagnetic to paramagnetic state occurs, at the point where the curve intersects $\textit{m}$=0. The empirical formula \cite{2011_A.B.Karki}, $\textit{B}_{c2}(T) = \textit{B}_{c2}(0) (1 - t^2) / (1 + t^2), \quad \text{where} \quad t = T / T_c$, is employed to fit the experimental data. The fit is depicted using the solid olive curve. From the GL fit, $\textit{B}_{c2}$(0) was determined as 7.38 T. 

Although the GL equations are expected to be \mbox{applicable} near the $\textit{T}_{\text{c}}$; in this study, it is observed that they accurately hold even to temperatures as low as $\sim$0.2 $\textit{T}_{\text{c}}$, which is characteristic of conventional BCS superconductors \cite{2022_Chuan-Yin_Xia}. The GL fit shown in Figure \ref{fig3} (b) shows a perfect linear relationship with temperature from $\textit{T}_{\text{c}}$ to $\sim$0.5 $\textit{T}_{\text{c}}$, which aligns with the behavior typical of superconductors with a high \mbox{Ginzburg-Landau parameter $\kappa$} \cite{2015_Y.Li}. 

Assuming that the breakdown of superconductivity at $\textit{B}_{\text{c2}}$ is through the orbital pair breaking effect \cite{2007_Y.Matsuda}, the superconducting coherence length $\xi$ and penetration depth $\lambda$ can be calculated using the GL equations, $\mu_0 \textit{H}_{\text{c2}} = \phi_0/2\pi\xi^2$ and $\mu_0 \textit{H}_{\text{c1}} = \phi_0/2\pi\lambda^2$, where $\phi_0$ denotes the flux quantum.

 The coherence length is estimated to be 6.7 nm and the penetration depth to be 22.5 nm, from the values of \mbox{$\textit{B}_{\text{c1}}$ = 0.65 T} and $\textit{B}_{\text{c2}}$ = 7.38 T. The Ginzburg-Landau parameter, $\kappa = \lambda/\xi$ = 3.35, is then calculated. It has been reported that $\textit{B}_{\text{c2}}$ = 25 T, $\xi$ = 4.2 nm \cite{1994_W.D.Wu} and \mbox{$\textit{B}_{\text{c2}}$ = 22 T,} $\xi$ = 3.8 nm, from the measurements of such parameters in bulk V$_3$Si single crystals \cite{1999_M.Yethiraj}. However values similar to the measurements in this study,  \mbox{$\textit{B}_{\text{c2}}$ = 8.1 T} and \mbox{$\xi$ = 6.27 nm,} were recently reported \cite{2023_Shuyue.Ding} from low temperature tunnelling experiments conducted on V$_3$Si (100) single crystals, and it was proposed that the increase in coherence length could be due to multi-band $\textit{s}$-wave superconductivity in V$_3$Si. The sheet resistance of the films was measured using a \mbox{4-probe} method at room temperature. Before annealing, the sheet resistance was 1.3 $\Omega/\square$ (ohm/square). After annealing, it increased to 3 $\Omega/\square$ for sample A and \mbox{5.8 $\Omega/\square$} for sample B, which shows that the sheet resistance increases with the increase in annealing temperature.

 The analysis of the results from the magnetization measurements in the prepared polycrystalline V$_3$Si film can be summarized as follows. It is a type-II, \mbox{high-$\textit{T}_{\text{c}}$}, high-$\kappa$ material in its superconducting state, known for its strong anisotropic properties \citep{christen1985}.
 Magnetic flux starts to penetrate the superconductor in distinct threads known as flux lines or vortices (fluxoids) when the superconductor is exposed to an external magnetic field that is stronger than the lower critical field $\textit{B}_{\text{c1}}$, but weaker than the upper critical field $\textit{B}_{\text{c2}}$. Each vortex carries one quantum of magnetic flux ($\phi_0$) and is surrounded by a circulating supercurrent. At the core of each vortex, the superconducting order parameter is significantly reduced, thereby effectively eliminating superconductivity locally within the core \citep{moshchalkov1996, maniv2001}. 

 The vortex pinning strength of a type II superconductor \citep{campbell1972} is an important parameter since it ensures that the zero electrical resistance is maintained under the influence of an external magnetic field. A high pinning efficiency restricts the movement of vortices when a current is applied, thereby enhancing the critical current density $\textit{J}_{\text{c}}$. Hence energy dissipation is minimized, there is stability in varying magnetic fields and superconducting properties are preserved.
 
 The fluxoid spacing, $a_0 = \sqrt{\frac{\Phi_0}{B}}$, where $\phi_0$ is the magnetic flux quantum, refers to the average distance between adjacent vortices in a type II superconductor in the mixed state. This spacing is dependent on the applied magnetic field strength. Thus, a large grain size compared to the small fluxoid spacing leads to lower pinning efficiency at high magnetic fields. However, the increase in the size of the grain boundaries compared to the grain size can likely improve or maintain pinning efficiency by compensating for the larger grain size. This is achieved by providing ample pinning sites across the material due to the extensive influence of the grain boundaries. 

The vortex pinning strength of these films can be estimated from the grain boundary density, which is inversely proportional to the grain size. For a given magnetic field, fluxoid spacing can be calculated. For example, at 4 T, for sample A, with a grain size of $\sim$33 nm, $a_0$ is calculated as 22 nm, which shows a slight decrease in pinning efficiency. Similarly, for sample B, with a grain size of $\sim$60 nm, the pinning efficiency further decreases at 4 T.

In reversible type-II superconductors, with an increase in the external magnetic field, the vortex density also rises, leading the vortices to organize into a highly ordered structure called the Abrikosov lattice \citep{abrikosov2004}, which minimizes the system's free energy. This lattice is generally characterized by an hexagonal configuration of vortices in high quality single crystal V$_3$Si samples \citep{Sosolik2003}. However, in polycrystalline V$_3$Si thin films, the Abrikosov lattice is disordered due to the presence of strong pinning centers such as grain boundaries, defects and impurities. Hence the typical hexagonal vortex arrangement is disrupted resulting in a metastable state with vortices trapped inhomogeneously within the sample, which causes irreversible behaviour in the magnetization of the superconductor. The Abrikosov unit cell, which is the smallest repetitive unit in this lattice, has an area that decreases as the magnetic field strength increases, indicating an inverse relationship between the two \citep{rosenstein2010}.

The vortex core size can be approximated based on the coherence length $\xi$, with the Ginzburg-Landau theory being applicable near the critical temperature, $\textit{T}_{\text{c}}$. The approximate area of the decay of the fluxoid core is given by $\pi\xi^2$, while $\pi\lambda^2$ represents the area through which the magnetic field penetrates without disrupting superconductivity, persisting up to an upper critical field \citep{de2018}. $\textit{B}_{\text{c2}}$ is estimated where the average spacing between the centres of adjacent vortices approximates the size of their cores. At this point, the superconducting regions connecting the fluxoids start to overlap significantly, and finally diminish. When reaching $\textit{B}_{\text{c2}}$, the vortex cores expand to such an extent that the continuous superconducting paths for current flow are disrupted, rendering the entire sample non-superconducting. This phenomenon supports the report of a smaller coherence length in bulk V$_3$Si single crystals.

The calculations in this study indicate a slightly large coherence length, which might be due to the increased size of the vortex core at low magnetic fields. This will result in weaker pinning strength and hence the degradation of the superconducting properties. The lower $\textit{B}_{\text{c2}}$ measured in this study compared to the bulk, is likely due to the larger coherence length calculated at zero field, since $\textit{B}_{\text{c2}}$ varies inversely with $\xi$. Vortex interactions are generally repulsive because their circulating currents generate magnetic fields that interact. The vortex density increases with magnetic field strength, decreasing the average distance between vortices and increasing the repulsive force between them.  The superconducting material reduces the overlap of the vortex cores by decreasing each vortex’s core area, in order to minimise the system’s energy. Hence, the core size reduces at low fields to sustain an energetically optimal vortex configuration.

In the dirty limit of the microscopic theory, it has been established that the size of the vortex core in a conventional s-wave superconductor reduces as the magnetic field intensifies \cite{1994_A.A.Golubov}. This effect was previously observed in V$_3$Si single crystals \cite{2003_Etienne_Boaknin}. Therefore, the extension of Ginzburg-Landau theory developed by Abrikosov that forms the foundation of type II superconductivity details the formation of the Abrikosov flux line lattice and its superconducting order parameters.

\section{Conclusions}
In conclusion, detailed investigations of the phase transformation during vanadium silicide thin-film formation are reported. A chemical model is proposed to evaluate the time dependency of the phase formation during the thin-film growth at high temperatures. Magnetization measurements have been performed on the annealed samples to assess superconductivity and the superconducting order parameters of the film were explored. This study provides a methodology for optimizing the growth conditions for superconducting thin films, which is an advancement towards developing future superconducting quantum devices that can operate well above liquid helium temperature. One of the future research avenues in this work is to optimize the chemical model for the growth of thin superconducting films. Another area worth exploring is the proximity effect induced by the substrate on the superconducting properties of these V$_3$Si films.
\section{Acknowledgments}
This work was performed in part at the Australian Nuclear Science and Technology Organization. We acknowledge the support of the Australian Centre for Neutron Scattering, ANSTO and the Australian Government through the National Collaborative Research Infrastructure Strategy (NCRIS), in supporting the neutron research infrastructure used in this work via ACNS proposal 17337. M. Bose is also supported in part by ARC grants (DP200103233 and CE170100012).

\bibliography{sample}

\begin{thebibliography}{65}%
\makeatletter
\providecommand \@ifxundefined [1]{%
 \@ifx{#1\undefined}
}%
\providecommand \@ifnum [1]{%
 \ifnum #1\expandafter \@firstoftwo
 \else \expandafter \@secondoftwo
 \fi
}%
\providecommand \@ifx [1]{%
 \ifx #1\expandafter \@firstoftwo
 \else \expandafter \@secondoftwo
 \fi
}%
\providecommand \natexlab [1]{#1}%
\providecommand \enquote  [1]{``#1''}%
\providecommand \bibnamefont  [1]{#1}%
\providecommand \bibfnamefont [1]{#1}%
\providecommand \citenamefont [1]{#1}%
\providecommand \href@noop [0]{\@secondoftwo}%
\providecommand \href [0]{\begingroup \@sanitize@url \@href}%
\providecommand \@href[1]{\@@startlink{#1}\@@href}%
\providecommand \@@href[1]{\endgroup#1\@@endlink}%
\providecommand \@sanitize@url [0]{\catcode `\\12\catcode `\$12\catcode `\&12\catcode `\#12\catcode `\^12\catcode `\_12\catcode `\%12\relax}%
\providecommand \@@startlink[1]{}%
\providecommand \@@endlink[0]{}%
\providecommand \url  [0]{\begingroup\@sanitize@url \@url }%
\providecommand \@url [1]{\endgroup\@href {#1}{\urlprefix }}%
\providecommand \urlprefix  [0]{URL }%
\providecommand \Eprint [0]{\href }%
\providecommand \doibase [0]{https://doi.org/}%
\providecommand \selectlanguage [0]{\@gobble}%
\providecommand \bibinfo  [0]{\@secondoftwo}%
\providecommand \bibfield  [0]{\@secondoftwo}%
\providecommand \translation [1]{[#1]}%
\providecommand \BibitemOpen [0]{}%
\providecommand \bibitemStop [0]{}%
\providecommand \bibitemNoStop [0]{.\EOS\space}%
\providecommand \EOS [0]{\spacefactor3000\relax}%
\providecommand \BibitemShut  [1]{\csname bibitem#1\endcsname}%
\let\auto@bib@innerbib\@empty
\bibitem [{\citenamefont {Kjaergaard}\ \emph {et~al.}(2020)\citenamefont {Kjaergaard}, \citenamefont {Schwartz}, \citenamefont {Braumüller}, \citenamefont {Krantz}, \citenamefont {Wang}, \citenamefont {Gustavsson},\ and\ \citenamefont {Oliver}}]{kjaergaard2020}%
  \BibitemOpen
  \bibfield  {author} {\bibinfo {author} {\bibfnamefont {M.}~\bibnamefont {Kjaergaard}}, \bibinfo {author} {\bibfnamefont {M.~E.}\ \bibnamefont {Schwartz}}, \bibinfo {author} {\bibfnamefont {J.}~\bibnamefont {Braumüller}}, \bibinfo {author} {\bibfnamefont {P.}~\bibnamefont {Krantz}}, \bibinfo {author} {\bibfnamefont {J.~I.-J.}\ \bibnamefont {Wang}}, \bibinfo {author} {\bibfnamefont {S.}~\bibnamefont {Gustavsson}},\ and\ \bibinfo {author} {\bibfnamefont {W.~D.}\ \bibnamefont {Oliver}},\ }\bibfield  {title} {\bibinfo {title} {Superconducting qubits: Current state of play},\ }\href {https://doi.org/https://doi.org/10.1146/annurev-conmatphys-031119-050605} {\bibfield  {journal} {\bibinfo  {journal} {Annual Review of Condensed Matter Physics}\ }\textbf {\bibinfo {volume} {11}},\ \bibinfo {pages} {369} (\bibinfo {year} {2020})}\BibitemShut {NoStop}%
\bibitem [{\citenamefont {Burkard}\ \emph {et~al.}(2020)\citenamefont {Burkard}, \citenamefont {Gullans}, \citenamefont {Mi},\ and\ \citenamefont {Petta}}]{burkard2020}%
  \BibitemOpen
  \bibfield  {author} {\bibinfo {author} {\bibfnamefont {G.}~\bibnamefont {Burkard}}, \bibinfo {author} {\bibfnamefont {M.}~\bibnamefont {Gullans}}, \bibinfo {author} {\bibfnamefont {X.}~\bibnamefont {Mi}},\ and\ \bibinfo {author} {\bibfnamefont {J.}~\bibnamefont {Petta}},\ }\bibfield  {title} {\bibinfo {title} {Superconductor–semiconductor hybrid-circuit quantum electrodynamics},\ }\href {https://doi.org/10.1038/s42254-019-0135-2} {\bibfield  {journal} {\bibinfo  {journal} {Nature Reviews Physics}\ }\textbf {\bibinfo {volume} {2}},\ \bibinfo {pages} {129} (\bibinfo {year} {2020})}\BibitemShut {NoStop}%
\bibitem [{\citenamefont {Bravyi}\ \emph {et~al.}(2022)\citenamefont {Bravyi}, \citenamefont {Dial}, \citenamefont {Gambetta}, \citenamefont {Gil},\ and\ \citenamefont {Nazario}}]{bravyi2022}%
  \BibitemOpen
  \bibfield  {author} {\bibinfo {author} {\bibfnamefont {S.}~\bibnamefont {Bravyi}}, \bibinfo {author} {\bibfnamefont {O.}~\bibnamefont {Dial}}, \bibinfo {author} {\bibfnamefont {J.~M.}\ \bibnamefont {Gambetta}}, \bibinfo {author} {\bibfnamefont {D.}~\bibnamefont {Gil}},\ and\ \bibinfo {author} {\bibfnamefont {Z.}~\bibnamefont {Nazario}},\ }\bibfield  {title} {\bibinfo {title} {{The future of quantum computing with superconducting qubits}},\ }\href {https://doi.org/10.1063/5.0082975} {\bibfield  {journal} {\bibinfo  {journal} {Journal of Applied Physics}\ }\textbf {\bibinfo {volume} {132}},\ \bibinfo {pages} {160902} (\bibinfo {year} {2022})}\BibitemShut {NoStop}%
\bibitem [{\citenamefont {Yang}\ \emph {et~al.}(2020)\citenamefont {Yang}, \citenamefont {Leon}, \citenamefont {Hwang}, \citenamefont {Saraiva}, \citenamefont {Tanttu}, \citenamefont {Huang}, \citenamefont {Camirand~Lemyre}, \citenamefont {Chan}, \citenamefont {Tan}, \citenamefont {Hudson} \emph {et~al.}}]{yang2020}%
  \BibitemOpen
  \bibfield  {author} {\bibinfo {author} {\bibfnamefont {C.~H.}\ \bibnamefont {Yang}}, \bibinfo {author} {\bibfnamefont {R.}~\bibnamefont {Leon}}, \bibinfo {author} {\bibfnamefont {J.}~\bibnamefont {Hwang}}, \bibinfo {author} {\bibfnamefont {A.}~\bibnamefont {Saraiva}}, \bibinfo {author} {\bibfnamefont {T.}~\bibnamefont {Tanttu}}, \bibinfo {author} {\bibfnamefont {W.}~\bibnamefont {Huang}}, \bibinfo {author} {\bibfnamefont {J.}~\bibnamefont {Camirand~Lemyre}}, \bibinfo {author} {\bibfnamefont {K.~W.}\ \bibnamefont {Chan}}, \bibinfo {author} {\bibfnamefont {K.}~\bibnamefont {Tan}}, \bibinfo {author} {\bibfnamefont {F.~E.}\ \bibnamefont {Hudson}}, \emph {et~al.},\ }\bibfield  {title} {\bibinfo {title} {Operation of a silicon quantum processor unit cell above one kelvin},\ }\href {https://doi.org/10.1038/s41586-020-2171-6} {\bibfield  {journal} {\bibinfo  {journal} {Nature}\ }\textbf {\bibinfo {volume} {580}},\ \bibinfo {pages} {350} (\bibinfo {year} {2020})}\BibitemShut {NoStop}%
\bibitem [{\citenamefont {Petit}\ \emph {et~al.}(2020)\citenamefont {Petit}, \citenamefont {Eenink}, \citenamefont {Russ}, \citenamefont {Lawrie}, \citenamefont {Hendrickx}, \citenamefont {Philips}, \citenamefont {Clarke}, \citenamefont {Vandersypen},\ and\ \citenamefont {Veldhorst}}]{petit2020}%
  \BibitemOpen
  \bibfield  {author} {\bibinfo {author} {\bibfnamefont {L.}~\bibnamefont {Petit}}, \bibinfo {author} {\bibfnamefont {H.}~\bibnamefont {Eenink}}, \bibinfo {author} {\bibfnamefont {M.}~\bibnamefont {Russ}}, \bibinfo {author} {\bibfnamefont {W.}~\bibnamefont {Lawrie}}, \bibinfo {author} {\bibfnamefont {N.}~\bibnamefont {Hendrickx}}, \bibinfo {author} {\bibfnamefont {S.}~\bibnamefont {Philips}}, \bibinfo {author} {\bibfnamefont {J.}~\bibnamefont {Clarke}}, \bibinfo {author} {\bibfnamefont {L.}~\bibnamefont {Vandersypen}},\ and\ \bibinfo {author} {\bibfnamefont {M.}~\bibnamefont {Veldhorst}},\ }\bibfield  {title} {\bibinfo {title} {Universal quantum logic in hot silicon qubits},\ }\href {https://doi.org/10.1038/s41586-020-2170-7} {\bibfield  {journal} {\bibinfo  {journal} {Nature}\ }\textbf {\bibinfo {volume} {580}},\ \bibinfo {pages} {355} (\bibinfo {year} {2020})}\BibitemShut {NoStop}%
\bibitem [{\citenamefont {Hardy}\ and\ \citenamefont {Hulm}(1953)}]{hardy1953}%
  \BibitemOpen
  \bibfield  {author} {\bibinfo {author} {\bibfnamefont {G.~F.}\ \bibnamefont {Hardy}}\ and\ \bibinfo {author} {\bibfnamefont {J.~K.}\ \bibnamefont {Hulm}},\ }\bibfield  {title} {\bibinfo {title} {Superconducting silicides and germanides},\ }\href {https://doi.org/10.1103/PhysRev.89.884} {\bibfield  {journal} {\bibinfo  {journal} {Phys. Rev.}\ }\textbf {\bibinfo {volume} {89}},\ \bibinfo {pages} {884} (\bibinfo {year} {1953})}\BibitemShut {NoStop}%
\bibitem [{\citenamefont {Hardy}\ and\ \citenamefont {Hulm}(1954)}]{hardy1954}%
  \BibitemOpen
  \bibfield  {author} {\bibinfo {author} {\bibfnamefont {G.~F.}\ \bibnamefont {Hardy}}\ and\ \bibinfo {author} {\bibfnamefont {J.~K.}\ \bibnamefont {Hulm}},\ }\bibfield  {title} {\bibinfo {title} {The superconductivity of some transition metal compounds},\ }\href {https://doi.org/10.1103/PhysRev.93.1004} {\bibfield  {journal} {\bibinfo  {journal} {Phys. Rev.}\ }\textbf {\bibinfo {volume} {93}},\ \bibinfo {pages} {1004} (\bibinfo {year} {1954})}\BibitemShut {NoStop}%
\bibitem [{\citenamefont {Saur}(1969)}]{Saur1969}%
  \BibitemOpen
  \bibfield  {author} {\bibinfo {author} {\bibfnamefont {E.}~\bibnamefont {Saur}},\ }\bibinfo {title} {Experiments on the critical behavior of type-ii superconductors in high magnetic fields},\ in\ \href {https://doi.org/10.1007/978-1-4899-5508-1_24} {\emph {\bibinfo {booktitle} {Physics of Solids in Intense Magnetic Fields: Lectures presented at the First Chania Conference held at Chania, Crete, July 16--29, 1967}}},\ \bibinfo {editor} {edited by\ \bibinfo {editor} {\bibfnamefont {E.~D.}\ \bibnamefont {Haidemenakis}}}\ (\bibinfo  {publisher} {Springer US},\ \bibinfo {address} {Boston, MA},\ \bibinfo {year} {1969})\ pp.\ \bibinfo {pages} {454--466}\BibitemShut {NoStop}%
\bibitem [{\citenamefont {Greiner}\ and\ \citenamefont {Mason}(1964)}]{Greiner1964}%
  \BibitemOpen
  \bibfield  {author} {\bibinfo {author} {\bibfnamefont {E.~S.}\ \bibnamefont {Greiner}}\ and\ \bibinfo {author} {\bibfnamefont {J.}~\bibnamefont {Mason}, \bibfnamefont {H.}},\ }\bibfield  {title} {\bibinfo {title} {Preparation of single crystals of $\textnormal{V}_3\textnormal{Si}$},\ }\href {https://doi.org/10.1063/1.1713170} {\bibfield  {journal} {\bibinfo  {journal} {Journal of Applied Physics}\ }\textbf {\bibinfo {volume} {35}},\ \bibinfo {pages} {3058} (\bibinfo {year} {1964})}\BibitemShut {NoStop}%
\bibitem [{\citenamefont {Shull}\ and\ \citenamefont {Wedgwood}(1966)}]{Shull1966}%
  \BibitemOpen
  \bibfield  {author} {\bibinfo {author} {\bibfnamefont {C.~G.}\ \bibnamefont {Shull}}\ and\ \bibinfo {author} {\bibfnamefont {F.~A.}\ \bibnamefont {Wedgwood}},\ }\bibfield  {title} {\bibinfo {title} {Neutron-diffraction studies of electron-spin pairing in superconducting $\textnormal{V}_3\textnormal{Si}$},\ }\href {https://doi.org/10.1103/PhysRevLett.16.513} {\bibfield  {journal} {\bibinfo  {journal} {Phys. Rev. Lett.}\ }\textbf {\bibinfo {volume} {16}},\ \bibinfo {pages} {513} (\bibinfo {year} {1966})}\BibitemShut {NoStop}%
\bibitem [{\citenamefont {Batterman}\ and\ \citenamefont {Barrett}(1966)}]{Batterman1966}%
  \BibitemOpen
  \bibfield  {author} {\bibinfo {author} {\bibfnamefont {B.~W.}\ \bibnamefont {Batterman}}\ and\ \bibinfo {author} {\bibfnamefont {C.~S.}\ \bibnamefont {Barrett}},\ }\bibfield  {title} {\bibinfo {title} {Low-temperature structural transformation in $\textnormal{V}_3\textnormal{Si}$},\ }\href {https://doi.org/10.1103/PhysRev.145.296} {\bibfield  {journal} {\bibinfo  {journal} {Phys. Rev.}\ }\textbf {\bibinfo {volume} {145}},\ \bibinfo {pages} {296} (\bibinfo {year} {1966})}\BibitemShut {NoStop}%
\bibitem [{\citenamefont {Testardi}\ and\ \citenamefont {Bateman}(1967)}]{Testardi1967}%
  \BibitemOpen
  \bibfield  {author} {\bibinfo {author} {\bibfnamefont {L.~R.}\ \bibnamefont {Testardi}}\ and\ \bibinfo {author} {\bibfnamefont {T.~B.}\ \bibnamefont {Bateman}},\ }\bibfield  {title} {\bibinfo {title} {Lattice instability of high-transition-temperature superconductors. {II}. single-crystal $\textnormal{V}_3\textnormal{Si}$ results},\ }\href {https://doi.org/10.1103/PhysRev.154.402} {\bibfield  {journal} {\bibinfo  {journal} {Phys. Rev.}\ }\textbf {\bibinfo {volume} {154}},\ \bibinfo {pages} {402} (\bibinfo {year} {1967})}\BibitemShut {NoStop}%
\bibitem [{\citenamefont {Kunzler}\ \emph {et~al.}(1966)\citenamefont {Kunzler}, \citenamefont {Maita}, \citenamefont {Levinstein},\ and\ \citenamefont {Ryder}}]{Kunzler1966}%
  \BibitemOpen
  \bibfield  {author} {\bibinfo {author} {\bibfnamefont {J.~E.}\ \bibnamefont {Kunzler}}, \bibinfo {author} {\bibfnamefont {J.~P.}\ \bibnamefont {Maita}}, \bibinfo {author} {\bibfnamefont {H.~J.}\ \bibnamefont {Levinstein}},\ and\ \bibinfo {author} {\bibfnamefont {E.~J.}\ \bibnamefont {Ryder}},\ }\bibfield  {title} {\bibinfo {title} {Pronounced change in the low-temperature heat capacity of $\textnormal{V}_3\textnormal{Si}$ with stress},\ }\href {https://doi.org/10.1103/PhysRev.143.390} {\bibfield  {journal} {\bibinfo  {journal} {Phys. Rev.}\ }\textbf {\bibinfo {volume} {143}},\ \bibinfo {pages} {390} (\bibinfo {year} {1966})}\BibitemShut {NoStop}%
\bibitem [{\citenamefont {King}\ \emph {et~al.}(1967)\citenamefont {King}, \citenamefont {Cocks},\ and\ \citenamefont {Pollock}}]{King1967}%
  \BibitemOpen
  \bibfield  {author} {\bibinfo {author} {\bibfnamefont {H.}~\bibnamefont {King}}, \bibinfo {author} {\bibfnamefont {F.}~\bibnamefont {Cocks}},\ and\ \bibinfo {author} {\bibfnamefont {J.}~\bibnamefont {Pollock}},\ }\bibfield  {title} {\bibinfo {title} {Further evidence of the low temperature phase transformation in $\textnormal{Nb}_3\textnormal{Sn}$ and $\textnormal{V}_3\textnormal{Si}$},\ }\href {https://doi.org/https://doi.org/10.1016/0375-9601(67)90109-0} {\bibfield  {journal} {\bibinfo  {journal} {Physics Letters A}\ }\textbf {\bibinfo {volume} {26}},\ \bibinfo {pages} {77} (\bibinfo {year} {1967})}\BibitemShut {NoStop}%
\bibitem [{\citenamefont {Mattheiss}\ and\ \citenamefont {Weber}(1982)}]{mattheiss1982}%
  \BibitemOpen
  \bibfield  {author} {\bibinfo {author} {\bibfnamefont {L.~F.}\ \bibnamefont {Mattheiss}}\ and\ \bibinfo {author} {\bibfnamefont {W.}~\bibnamefont {Weber}},\ }\bibfield  {title} {\bibinfo {title} {Electronic structure of cubic $\textnormal{Nb}_3\textnormal{Sn}$ and $\textnormal{V}_3\textnormal{Si}$},\ }\href {https://doi.org/10.1103/PhysRevB.25.2248} {\bibfield  {journal} {\bibinfo  {journal} {Phys. Rev. B}\ }\textbf {\bibinfo {volume} {25}},\ \bibinfo {pages} {2248} (\bibinfo {year} {1982})}\BibitemShut {NoStop}%
\bibitem [{\citenamefont {Bisi}\ and\ \citenamefont {Chiao}(1982)}]{bisi1982}%
  \BibitemOpen
  \bibfield  {author} {\bibinfo {author} {\bibfnamefont {O.}~\bibnamefont {Bisi}}\ and\ \bibinfo {author} {\bibfnamefont {L.~W.}\ \bibnamefont {Chiao}},\ }\bibfield  {title} {\bibinfo {title} {Electronic structure of vanadium silicides},\ }\href {https://doi.org/10.1103/PhysRevB.25.4943} {\bibfield  {journal} {\bibinfo  {journal} {Phys. Rev. B}\ }\textbf {\bibinfo {volume} {25}},\ \bibinfo {pages} {4943} (\bibinfo {year} {1982})}\BibitemShut {NoStop}%
\bibitem [{\citenamefont {Testardi}(1975)}]{Testardi1975}%
  \BibitemOpen
  \bibfield  {author} {\bibinfo {author} {\bibfnamefont {L.~R.}\ \bibnamefont {Testardi}},\ }\bibfield  {title} {\bibinfo {title} {Structural instability and superconductivity in {A}-15 compounds},\ }\href {https://doi.org/10.1103/RevModPhys.47.637} {\bibfield  {journal} {\bibinfo  {journal} {Rev. Mod. Phys.}\ }\textbf {\bibinfo {volume} {47}},\ \bibinfo {pages} {637} (\bibinfo {year} {1975})}\BibitemShut {NoStop}%
\bibitem [{\citenamefont {Dierker}\ \emph {et~al.}(1983)\citenamefont {Dierker}, \citenamefont {Klein}, \citenamefont {Webb},\ and\ \citenamefont {Fisk}}]{Dierker1983}%
  \BibitemOpen
  \bibfield  {author} {\bibinfo {author} {\bibfnamefont {S.~B.}\ \bibnamefont {Dierker}}, \bibinfo {author} {\bibfnamefont {M.~V.}\ \bibnamefont {Klein}}, \bibinfo {author} {\bibfnamefont {G.~W.}\ \bibnamefont {Webb}},\ and\ \bibinfo {author} {\bibfnamefont {Z.}~\bibnamefont {Fisk}},\ }\bibfield  {title} {\bibinfo {title} {Electronic raman scattering by superconducting-gap excitations in $\textnormal{Nb}_3\textnormal{Sn}$ and $\textnormal{V}_3\textnormal{Si}$},\ }\href {https://doi.org/10.1103/PhysRevLett.50.853} {\bibfield  {journal} {\bibinfo  {journal} {Phys. Rev. Lett.}\ }\textbf {\bibinfo {volume} {50}},\ \bibinfo {pages} {853} (\bibinfo {year} {1983})}\BibitemShut {NoStop}%
\bibitem [{\citenamefont {Mitrovi\ifmmode~\acute{c}\else \'{c}\fi{}}\ and\ \citenamefont {Carbotte}(1986)}]{Mitrovi1986}%
  \BibitemOpen
  \bibfield  {author} {\bibinfo {author} {\bibfnamefont {B.}~\bibnamefont {Mitrovi\ifmmode~\acute{c}\else \'{c}\fi{}}}\ and\ \bibinfo {author} {\bibfnamefont {J.~P.}\ \bibnamefont {Carbotte}},\ }\bibfield  {title} {\bibinfo {title} {Thermodynamic and far-infrared properties of $\textnormal{V}_3\textnormal{Si}$ calculated from tunneling results for the eliashberg function ${{\alpha^{2}F}}$ and coulomb pseudopotential parameter ${{\mu^*}}$},\ }\href {https://doi.org/10.1103/PhysRevB.33.591} {\bibfield  {journal} {\bibinfo  {journal} {Phys. Rev. B}\ }\textbf {\bibinfo {volume} {33}},\ \bibinfo {pages} {591} (\bibinfo {year} {1986})}\BibitemShut {NoStop}%
\bibitem [{\citenamefont {Hirota}\ \emph {et~al.}(1995)\citenamefont {Hirota}, \citenamefont {Rebelsky},\ and\ \citenamefont {Shirane}}]{Hirota1995}%
  \BibitemOpen
  \bibfield  {author} {\bibinfo {author} {\bibfnamefont {K.}~\bibnamefont {Hirota}}, \bibinfo {author} {\bibfnamefont {L.}~\bibnamefont {Rebelsky}},\ and\ \bibinfo {author} {\bibfnamefont {G.}~\bibnamefont {Shirane}},\ }\bibfield  {title} {\bibinfo {title} {X-ray-scattering study of the cubic-to-tetragonal transition and its precursive phenomenon in $\textnormal{V}_3\textnormal{Si}$},\ }\href {https://doi.org/10.1103/PhysRevB.51.11325} {\bibfield  {journal} {\bibinfo  {journal} {Phys. Rev. B}\ }\textbf {\bibinfo {volume} {51}},\ \bibinfo {pages} {11325} (\bibinfo {year} {1995})}\BibitemShut {NoStop}%
\bibitem [{\citenamefont {Suga}\ \emph {et~al.}(2012)\citenamefont {Suga}, \citenamefont {Itoda}, \citenamefont {Sekiyama}, \citenamefont {Fujiwara}, \citenamefont {Komori}, \citenamefont {Imada}, \citenamefont {Yabashi}, \citenamefont {Tamasaku}, \citenamefont {Higashiya}, \citenamefont {Ishikawa}, \citenamefont {Shang},\ and\ \citenamefont {Fujikawa}}]{Suga2012}%
  \BibitemOpen
  \bibfield  {author} {\bibinfo {author} {\bibfnamefont {S.}~\bibnamefont {Suga}}, \bibinfo {author} {\bibfnamefont {S.}~\bibnamefont {Itoda}}, \bibinfo {author} {\bibfnamefont {A.}~\bibnamefont {Sekiyama}}, \bibinfo {author} {\bibfnamefont {H.}~\bibnamefont {Fujiwara}}, \bibinfo {author} {\bibfnamefont {S.}~\bibnamefont {Komori}}, \bibinfo {author} {\bibfnamefont {S.}~\bibnamefont {Imada}}, \bibinfo {author} {\bibfnamefont {M.}~\bibnamefont {Yabashi}}, \bibinfo {author} {\bibfnamefont {K.}~\bibnamefont {Tamasaku}}, \bibinfo {author} {\bibfnamefont {A.}~\bibnamefont {Higashiya}}, \bibinfo {author} {\bibfnamefont {T.}~\bibnamefont {Ishikawa}}, \bibinfo {author} {\bibfnamefont {M.}~\bibnamefont {Shang}},\ and\ \bibinfo {author} {\bibfnamefont {T.}~\bibnamefont {Fujikawa}},\ }\bibfield  {title} {\bibinfo {title} {Recoil effects for valence and core photoelectrons in $\textnormal{V}_3\textnormal{Si}$},\ }\href {https://doi.org/10.1103/PhysRevB.86.035146} {\bibfield  {journal} {\bibinfo  {journal} {Phys. Rev. B}\
  }\textbf {\bibinfo {volume} {86}},\ \bibinfo {pages} {035146} (\bibinfo {year} {2012})}\BibitemShut {NoStop}%
\bibitem [{\citenamefont {Howard}\ \emph {et~al.}(2023)\citenamefont {Howard}, \citenamefont {Liepe},\ and\ \citenamefont {Sun}}]{Howard2023}%
  \BibitemOpen
  \bibfield  {author} {\bibinfo {author} {\bibfnamefont {K.}~\bibnamefont {Howard}}, \bibinfo {author} {\bibfnamefont {M.~U.}\ \bibnamefont {Liepe}},\ and\ \bibinfo {author} {\bibfnamefont {Z.}~\bibnamefont {Sun}},\ }\bibfield  {title} {\bibinfo {title} {{Thermal annealing of DC sputtered Nb$_3$Sn and V$_3$Si thin films for superconducting radio-frequency cavities}},\ }\href {https://doi.org/10.1063/5.0185404} {\bibfield  {journal} {\bibinfo  {journal} {Journal of Applied Physics}\ }\textbf {\bibinfo {volume} {134}},\ \bibinfo {pages} {225301} (\bibinfo {year} {2023})}\BibitemShut {NoStop}%
\bibitem [{\citenamefont {Zhang}\ \emph {et~al.}(2021)\citenamefont {Zhang}, \citenamefont {Bollinger}, \citenamefont {Li}, \citenamefont {Kisslinger}, \citenamefont {Tong}, \citenamefont {Liu},\ and\ \citenamefont {Black}}]{zhang2021}%
  \BibitemOpen
  \bibfield  {author} {\bibinfo {author} {\bibfnamefont {W.}~\bibnamefont {Zhang}}, \bibinfo {author} {\bibfnamefont {A.~T.}\ \bibnamefont {Bollinger}}, \bibinfo {author} {\bibfnamefont {R.}~\bibnamefont {Li}}, \bibinfo {author} {\bibfnamefont {K.}~\bibnamefont {Kisslinger}}, \bibinfo {author} {\bibfnamefont {X.}~\bibnamefont {Tong}}, \bibinfo {author} {\bibfnamefont {M.}~\bibnamefont {Liu}},\ and\ \bibinfo {author} {\bibfnamefont {C.~T.}\ \bibnamefont {Black}},\ }\bibfield  {title} {\bibinfo {title} {Thin-film synthesis of superconductor-on-insulator {A}15 vanadium silicide},\ }\href {https://doi.org/10.1038/s41598-021-82046-1} {\bibfield  {journal} {\bibinfo  {journal} {Scientific Reports}\ }\textbf {\bibinfo {volume} {11}},\ \bibinfo {pages} {2358} (\bibinfo {year} {2021})}\BibitemShut {NoStop}%
\bibitem [{\citenamefont {Bangert}\ \emph {et~al.}(1985)\citenamefont {Bangert}, \citenamefont {Geerk},\ and\ \citenamefont {Schweiss}}]{Bangert1985}%
  \BibitemOpen
  \bibfield  {author} {\bibinfo {author} {\bibfnamefont {W.}~\bibnamefont {Bangert}}, \bibinfo {author} {\bibfnamefont {J.}~\bibnamefont {Geerk}},\ and\ \bibinfo {author} {\bibfnamefont {P.}~\bibnamefont {Schweiss}},\ }\bibfield  {title} {\bibinfo {title} {Tunneling and neutron scattering experiments on {A}15 $\textnormal{V}_3\textnormal{Si}$},\ }\href {https://doi.org/10.1103/PhysRevB.31.6066} {\bibfield  {journal} {\bibinfo  {journal} {Phys. Rev. B}\ }\textbf {\bibinfo {volume} {31}},\ \bibinfo {pages} {6066} (\bibinfo {year} {1985})}\BibitemShut {NoStop}%
\bibitem [{\citenamefont {Borghesi}\ \emph {et~al.}(1989)\citenamefont {Borghesi}, \citenamefont {Piaggi}, \citenamefont {Guizzetti}, \citenamefont {Nava},\ and\ \citenamefont {Bacchetta}}]{Borghesi1989}%
  \BibitemOpen
  \bibfield  {author} {\bibinfo {author} {\bibfnamefont {A.}~\bibnamefont {Borghesi}}, \bibinfo {author} {\bibfnamefont {A.}~\bibnamefont {Piaggi}}, \bibinfo {author} {\bibfnamefont {G.}~\bibnamefont {Guizzetti}}, \bibinfo {author} {\bibfnamefont {F.}~\bibnamefont {Nava}},\ and\ \bibinfo {author} {\bibfnamefont {M.}~\bibnamefont {Bacchetta}},\ }\bibfield  {title} {\bibinfo {title} {Optical properties of vanadium silicide polycrystalline films},\ }\href {https://doi.org/10.1103/PhysRevB.40.3249} {\bibfield  {journal} {\bibinfo  {journal} {Phys. Rev. B}\ }\textbf {\bibinfo {volume} {40}},\ \bibinfo {pages} {3249} (\bibinfo {year} {1989})}\BibitemShut {NoStop}%
\bibitem [{\citenamefont {Oya}\ \emph {et~al.}(1982)\citenamefont {Oya}, \citenamefont {Inabe}, \citenamefont {Onodera},\ and\ \citenamefont {Sawada}}]{Oya1982}%
  \BibitemOpen
  \bibfield  {author} {\bibinfo {author} {\bibfnamefont {G.}~\bibnamefont {Oya}}, \bibinfo {author} {\bibfnamefont {H.}~\bibnamefont {Inabe}}, \bibinfo {author} {\bibfnamefont {Y.}~\bibnamefont {Onodera}},\ and\ \bibinfo {author} {\bibfnamefont {Y.}~\bibnamefont {Sawada}},\ }\bibfield  {title} {\bibinfo {title} {{Superconducting transition temperatures of thin V$_3$Si layers formed by the interaction of V films with thinly oxidized Si wafers}},\ }\href {https://doi.org/10.1063/1.330524} {\bibfield  {journal} {\bibinfo  {journal} {Journal of Applied Physics}\ }\textbf {\bibinfo {volume} {53}},\ \bibinfo {pages} {1115} (\bibinfo {year} {1982})}\BibitemShut {NoStop}%
\bibitem [{\citenamefont {Psaras}\ \emph {et~al.}(1984)\citenamefont {Psaras}, \citenamefont {Eizenberg},\ and\ \citenamefont {Tu}}]{Psaras1984}%
  \BibitemOpen
  \bibfield  {author} {\bibinfo {author} {\bibfnamefont {P.~A.}\ \bibnamefont {Psaras}}, \bibinfo {author} {\bibfnamefont {M.}~\bibnamefont {Eizenberg}},\ and\ \bibinfo {author} {\bibfnamefont {K.~N.}\ \bibnamefont {Tu}},\ }\bibfield  {title} {\bibinfo {title} {{Sequential silicide formation between vanadium and amorphous silicon thin‐film bilayers}},\ }\href {https://doi.org/10.1063/1.333910} {\bibfield  {journal} {\bibinfo  {journal} {Journal of Applied Physics}\ }\textbf {\bibinfo {volume} {56}},\ \bibinfo {pages} {3439} (\bibinfo {year} {1984})}\BibitemShut {NoStop}%
\bibitem [{\citenamefont {Derunova}\ \emph {et~al.}(2019)\citenamefont {Derunova}, \citenamefont {Sun}, \citenamefont {Felser}, \citenamefont {Parkin}, \citenamefont {Yan},\ and\ \citenamefont {Ali}}]{derunova2019}%
  \BibitemOpen
  \bibfield  {author} {\bibinfo {author} {\bibfnamefont {E.}~\bibnamefont {Derunova}}, \bibinfo {author} {\bibfnamefont {Y.}~\bibnamefont {Sun}}, \bibinfo {author} {\bibfnamefont {C.}~\bibnamefont {Felser}}, \bibinfo {author} {\bibfnamefont {S.~S.~P.}\ \bibnamefont {Parkin}}, \bibinfo {author} {\bibfnamefont {B.}~\bibnamefont {Yan}},\ and\ \bibinfo {author} {\bibfnamefont {M.~N.}\ \bibnamefont {Ali}},\ }\bibfield  {title} {\bibinfo {title} {Giant intrinsic spin hall effect in $\textnormal{W}_3\textnormal{Ta}$ and other {A}15 superconductors},\ }\href {https://doi.org/10.1126/sciadv.aav8575} {\bibfield  {journal} {\bibinfo  {journal} {Science Advances}\ }\textbf {\bibinfo {volume} {5}},\ \bibinfo {pages} {eaav8575} (\bibinfo {year} {2019})}\BibitemShut {NoStop}%
\bibitem [{\citenamefont {Vethaak}\ \emph {et~al.}(2021)\citenamefont {Vethaak}, \citenamefont {Gustavo}, \citenamefont {Farjot}, \citenamefont {Kubart}, \citenamefont {Gergaud}, \citenamefont {Zhang}, \citenamefont {Lefloch},\ and\ \citenamefont {Nemouchi}}]{vethaak2021}%
  \BibitemOpen
  \bibfield  {author} {\bibinfo {author} {\bibfnamefont {T.}~\bibnamefont {Vethaak}}, \bibinfo {author} {\bibfnamefont {F.}~\bibnamefont {Gustavo}}, \bibinfo {author} {\bibfnamefont {T.}~\bibnamefont {Farjot}}, \bibinfo {author} {\bibfnamefont {T.}~\bibnamefont {Kubart}}, \bibinfo {author} {\bibfnamefont {P.}~\bibnamefont {Gergaud}}, \bibinfo {author} {\bibfnamefont {S.-L.}\ \bibnamefont {Zhang}}, \bibinfo {author} {\bibfnamefont {F.}~\bibnamefont {Lefloch}},\ and\ \bibinfo {author} {\bibfnamefont {F.}~\bibnamefont {Nemouchi}},\ }\bibfield  {title} {\bibinfo {title} {Superconducting v3si for quantum circuit applications},\ }\href {https://doi.org/https://doi.org/10.1016/j.mee.2021.111570} {\bibfield  {journal} {\bibinfo  {journal} {Microelectronic Engineering}\ }\textbf {\bibinfo {volume} {244-246}},\ \bibinfo {pages} {111570} (\bibinfo {year} {2021})}\BibitemShut {NoStop}%
\bibitem [{\citenamefont {Torikai}(2011)}]{imae2011neutrons}%
  \BibitemOpen
  \bibfield  {author} {\bibinfo {author} {\bibfnamefont {N.}~\bibnamefont {Torikai}},\ }\bibinfo {title} {Neutron reflectometry},\ in\ \href {https://doi.org/https://doi.org/10.1002/9780470933886.ch5} {\emph {\bibinfo {booktitle} {Neutrons in Soft Matter}}}\ (\bibinfo  {publisher} {John Wiley \& Sons, Ltd},\ \bibinfo {year} {2011})\ Chap.\ \bibinfo {chapter} {II.2}, pp.\ \bibinfo {pages} {115--145}\BibitemShut {NoStop}%
\bibitem [{\citenamefont {Shirane}\ \emph {et~al.}(1971)\citenamefont {Shirane}, \citenamefont {Axe},\ and\ \citenamefont {Birgeneau}}]{Shirane1971}%
  \BibitemOpen
  \bibfield  {author} {\bibinfo {author} {\bibfnamefont {G.}~\bibnamefont {Shirane}}, \bibinfo {author} {\bibfnamefont {J.}~\bibnamefont {Axe}},\ and\ \bibinfo {author} {\bibfnamefont {R.}~\bibnamefont {Birgeneau}},\ }\bibfield  {title} {\bibinfo {title} {Neutron scattering study of the lattice dynamical phase transition in $\textnormal{V}_3\textnormal{Si}$},\ }\href {https://doi.org/https://doi.org/10.1016/0038-1098(71)90530-8} {\bibfield  {journal} {\bibinfo  {journal} {Solid State Communications}\ }\textbf {\bibinfo {volume} {9}},\ \bibinfo {pages} {397} (\bibinfo {year} {1971})}\BibitemShut {NoStop}%
\bibitem [{\citenamefont {Knapp}\ \emph {et~al.}(1976)\citenamefont {Knapp}, \citenamefont {Bader},\ and\ \citenamefont {Fisk}}]{knapp1976}%
  \BibitemOpen
  \bibfield  {author} {\bibinfo {author} {\bibfnamefont {G.~S.}\ \bibnamefont {Knapp}}, \bibinfo {author} {\bibfnamefont {S.~D.}\ \bibnamefont {Bader}},\ and\ \bibinfo {author} {\bibfnamefont {Z.}~\bibnamefont {Fisk}},\ }\bibfield  {title} {\bibinfo {title} {Phonon properties of ${A}{-}15$ superconductors obtained from heat-capacity measurements},\ }\href {https://doi.org/10.1103/PhysRevB.13.3783} {\bibfield  {journal} {\bibinfo  {journal} {Phys. Rev. B}\ }\textbf {\bibinfo {volume} {13}},\ \bibinfo {pages} {3783} (\bibinfo {year} {1976})}\BibitemShut {NoStop}%
\bibitem [{\citenamefont {Yethiraj}\ \emph {et~al.}(2005)\citenamefont {Yethiraj}, \citenamefont {Christen}, \citenamefont {Gapud}, \citenamefont {Paul}, \citenamefont {Crowe}, \citenamefont {Dewhurst}, \citenamefont {Cubitt}, \citenamefont {Porcar},\ and\ \citenamefont {Gurevich}}]{yethiraj2005}%
  \BibitemOpen
  \bibfield  {author} {\bibinfo {author} {\bibfnamefont {M.}~\bibnamefont {Yethiraj}}, \bibinfo {author} {\bibfnamefont {D.~K.}\ \bibnamefont {Christen}}, \bibinfo {author} {\bibfnamefont {A.~A.}\ \bibnamefont {Gapud}}, \bibinfo {author} {\bibfnamefont {D.~M.}\ \bibnamefont {Paul}}, \bibinfo {author} {\bibfnamefont {S.~J.}\ \bibnamefont {Crowe}}, \bibinfo {author} {\bibfnamefont {C.~D.}\ \bibnamefont {Dewhurst}}, \bibinfo {author} {\bibfnamefont {R.}~\bibnamefont {Cubitt}}, \bibinfo {author} {\bibfnamefont {L.}~\bibnamefont {Porcar}},\ and\ \bibinfo {author} {\bibfnamefont {A.}~\bibnamefont {Gurevich}},\ }\bibfield  {title} {\bibinfo {title} {Temperature and field dependence of the flux-line-lattice symmetry in $\textnormal{V}_3\textnormal{Si}$},\ }\href {https://doi.org/10.1103/PhysRevB.72.060504} {\bibfield  {journal} {\bibinfo  {journal} {Phys. Rev. B}\ }\textbf {\bibinfo {volume} {72}},\ \bibinfo {pages} {060504} (\bibinfo {year} {2005})}\BibitemShut {NoStop}%
\bibitem [{\citenamefont {Le~Brun}\ \emph {et~al.}(2023)\citenamefont {Le~Brun}, \citenamefont {Huang}, \citenamefont {Pullen}, \citenamefont {Nelson}, \citenamefont {Spedding},\ and\ \citenamefont {Holt}}]{le2023spatz}%
  \BibitemOpen
  \bibfield  {author} {\bibinfo {author} {\bibfnamefont {A.~P.}\ \bibnamefont {Le~Brun}}, \bibinfo {author} {\bibfnamefont {T.-Y.}\ \bibnamefont {Huang}}, \bibinfo {author} {\bibfnamefont {S.}~\bibnamefont {Pullen}}, \bibinfo {author} {\bibfnamefont {A.~R.~J.}\ \bibnamefont {Nelson}}, \bibinfo {author} {\bibfnamefont {J.}~\bibnamefont {Spedding}},\ and\ \bibinfo {author} {\bibfnamefont {S.~A.}\ \bibnamefont {Holt}},\ }\bibfield  {title} {\bibinfo {title} {{Spatz: the time-of-flight neutron reflectometer with vertical sample geometry at the OPAL research reactor}},\ }\href {https://doi.org/10.1107/S160057672201086X} {\bibfield  {journal} {\bibinfo  {journal} {Journal of Applied Crystallography}\ }\textbf {\bibinfo {volume} {56}},\ \bibinfo {pages} {18} (\bibinfo {year} {2023})}\BibitemShut {NoStop}%
\bibitem [{\citenamefont {Nelson}\ and\ \citenamefont {Prescott}(2019)}]{nelson2019refnx}%
  \BibitemOpen
  \bibfield  {author} {\bibinfo {author} {\bibfnamefont {A.~R.~J.}\ \bibnamefont {Nelson}}\ and\ \bibinfo {author} {\bibfnamefont {S.~W.}\ \bibnamefont {Prescott}},\ }\bibfield  {title} {\bibinfo {title} {{{\it refnx}: neutron and X-ray reflectometry analysis in Python}},\ }\href {https://doi.org/10.1107/S1600576718017296} {\bibfield  {journal} {\bibinfo  {journal} {Journal of Applied Crystallography}\ }\textbf {\bibinfo {volume} {52}},\ \bibinfo {pages} {193} (\bibinfo {year} {2019})}\BibitemShut {NoStop}%
\bibitem [{\citenamefont {Holt}\ \emph {et~al.}(2022)\citenamefont {Holt}, \citenamefont {Oliver},\ and\ \citenamefont {Nelson}}]{holt2022using}%
  \BibitemOpen
  \bibfield  {author} {\bibinfo {author} {\bibfnamefont {S.~A.}\ \bibnamefont {Holt}}, \bibinfo {author} {\bibfnamefont {T.~E.}\ \bibnamefont {Oliver}},\ and\ \bibinfo {author} {\bibfnamefont {A.~R.~J.}\ \bibnamefont {Nelson}},\ }\bibinfo {title} {Using refnx to model neutron reflectometry data from phospholipid bilayers},\ in\ \href {https://doi.org/10.1007/978-1-0716-1843-1_15} {\emph {\bibinfo {booktitle} {Membrane Lipids: Methods and Protocols}}},\ \bibinfo {editor} {edited by\ \bibinfo {editor} {\bibfnamefont {C.~G.}\ \bibnamefont {Cranfield}}}\ (\bibinfo  {publisher} {Springer US},\ \bibinfo {address} {New York, NY},\ \bibinfo {year} {2022})\ pp.\ \bibinfo {pages} {179--197}\BibitemShut {NoStop}%
\bibitem [{\citenamefont {Goodway}\ \emph {et~al.}(2019)\citenamefont {Goodway}, \citenamefont {McIntyre}, \citenamefont {Sears}, \citenamefont {Belkhier}, \citenamefont {Burgess}, \citenamefont {Kirichek}, \citenamefont {Leli{\`e}vre-Berna}, \citenamefont {Marchal}, \citenamefont {Turc},\ and\ \citenamefont {Wakefield}}]{goodway2019fast}%
  \BibitemOpen
  \bibfield  {author} {\bibinfo {author} {\bibfnamefont {C.}~\bibnamefont {Goodway}}, \bibinfo {author} {\bibfnamefont {P.}~\bibnamefont {McIntyre}}, \bibinfo {author} {\bibfnamefont {A.}~\bibnamefont {Sears}}, \bibinfo {author} {\bibfnamefont {N.}~\bibnamefont {Belkhier}}, \bibinfo {author} {\bibfnamefont {G.}~\bibnamefont {Burgess}}, \bibinfo {author} {\bibfnamefont {O.}~\bibnamefont {Kirichek}}, \bibinfo {author} {\bibfnamefont {E.}~\bibnamefont {Leli{\`e}vre-Berna}}, \bibinfo {author} {\bibfnamefont {F.}~\bibnamefont {Marchal}}, \bibinfo {author} {\bibfnamefont {S.}~\bibnamefont {Turc}},\ and\ \bibinfo {author} {\bibfnamefont {S.}~\bibnamefont {Wakefield}},\ }\bibfield  {title} {\bibinfo {title} {A fast-cooling mode for blue series furnaces},\ }\href {https://doi.org/10.3233/JNR-190128} {\bibfield  {journal} {\bibinfo  {journal} {Journal of Neutron Research}\ }\textbf {\bibinfo {volume} {21}},\ \bibinfo {pages} {137} (\bibinfo {year} {2019})}\BibitemShut {NoStop}%
\bibitem [{\citenamefont {Livingston}(1977)}]{livingston1977}%
  \BibitemOpen
  \bibfield  {author} {\bibinfo {author} {\bibfnamefont {J.~D.}\ \bibnamefont {Livingston}},\ }\bibfield  {title} {\bibinfo {title} {Grain size in {A}-15 reaction layers},\ }\href {https://doi.org/https://doi.org/10.1002/pssa.2210440131} {\bibfield  {journal} {\bibinfo  {journal} {physica status solidi (a)}\ }\textbf {\bibinfo {volume} {44}},\ \bibinfo {pages} {295} (\bibinfo {year} {1977})}\BibitemShut {NoStop}%
\bibitem [{\citenamefont {Godeke}\ \emph {et~al.}(2005)\citenamefont {Godeke}, \citenamefont {Jewell}, \citenamefont {Fischer}, \citenamefont {Squitieri}, \citenamefont {Lee},\ and\ \citenamefont {Larbalestier}}]{godeke2005}%
  \BibitemOpen
  \bibfield  {author} {\bibinfo {author} {\bibfnamefont {A.}~\bibnamefont {Godeke}}, \bibinfo {author} {\bibfnamefont {M.~C.}\ \bibnamefont {Jewell}}, \bibinfo {author} {\bibfnamefont {C.~M.}\ \bibnamefont {Fischer}}, \bibinfo {author} {\bibfnamefont {A.~A.}\ \bibnamefont {Squitieri}}, \bibinfo {author} {\bibfnamefont {P.~J.}\ \bibnamefont {Lee}},\ and\ \bibinfo {author} {\bibfnamefont {D.~C.}\ \bibnamefont {Larbalestier}},\ }\bibfield  {title} {\bibinfo {title} {{The upper critical field of filamentary Nb$_3$Sn conductors}},\ }\href {https://doi.org/10.1063/1.1890447} {\bibfield  {journal} {\bibinfo  {journal} {Journal of Applied Physics}\ }\textbf {\bibinfo {volume} {97}},\ \bibinfo {pages} {093909} (\bibinfo {year} {2005})}\BibitemShut {NoStop}%
\bibitem [{\citenamefont {Goldfarb}\ \emph {et~al.}(1991)\citenamefont {Goldfarb}, \citenamefont {Lelental},\ and\ \citenamefont {Thompson}}]{Goldfarb1991}%
  \BibitemOpen
  \bibfield  {author} {\bibinfo {author} {\bibfnamefont {R.~B.}\ \bibnamefont {Goldfarb}}, \bibinfo {author} {\bibfnamefont {M.}~\bibnamefont {Lelental}},\ and\ \bibinfo {author} {\bibfnamefont {C.~A.}\ \bibnamefont {Thompson}},\ }\bibinfo {title} {Alternating-field susceptometry and magnetic susceptibility of superconductors},\ in\ \href {https://doi.org/10.1007/978-1-4899-2379-0_3} {\emph {\bibinfo {booktitle} {Magnetic Susceptibility of Superconductors and Other Spin Systems}}},\ \bibinfo {editor} {edited by\ \bibinfo {editor} {\bibfnamefont {R.~A.}\ \bibnamefont {Hein}}, \bibinfo {editor} {\bibfnamefont {T.~L.}\ \bibnamefont {Francavilla}},\ and\ \bibinfo {editor} {\bibfnamefont {D.~H.}\ \bibnamefont {Liebenberg}}}\ (\bibinfo  {publisher} {Springer US},\ \bibinfo {address} {Boston, MA},\ \bibinfo {year} {1991})\ pp.\ \bibinfo {pages} {49--80}\BibitemShut {NoStop}%
\bibitem [{\citenamefont {Hauser}\ and\ \citenamefont {Theuerer}(1963)}]{Hauser1963}%
  \BibitemOpen
  \bibfield  {author} {\bibinfo {author} {\bibfnamefont {J.~J.}\ \bibnamefont {Hauser}}\ and\ \bibinfo {author} {\bibfnamefont {H.~C.}\ \bibnamefont {Theuerer}},\ }\bibfield  {title} {\bibinfo {title} {Evidence for filaments in ${\mathrm{v}}_{3}$si},\ }\href {https://doi.org/10.1103/PhysRev.129.103} {\bibfield  {journal} {\bibinfo  {journal} {Phys. Rev.}\ }\textbf {\bibinfo {volume} {129}},\ \bibinfo {pages} {103} (\bibinfo {year} {1963})}\BibitemShut {NoStop}%
\bibitem [{\citenamefont {Croke}\ \emph {et~al.}(1988)\citenamefont {Croke}, \citenamefont {Hauenstein},\ and\ \citenamefont {McGill}}]{ETCroke1988}%
  \BibitemOpen
  \bibfield  {author} {\bibinfo {author} {\bibfnamefont {E.~T.}\ \bibnamefont {Croke}}, \bibinfo {author} {\bibfnamefont {R.~J.}\ \bibnamefont {Hauenstein}},\ and\ \bibinfo {author} {\bibfnamefont {T.~C.}\ \bibnamefont {McGill}},\ }\bibfield  {title} {\bibinfo {title} {{Growth of superconducting V3Si on Si by molecular beam epitaxial techniques}},\ }\href@noop {} {\bibfield  {journal} {\bibinfo  {journal} {Applied Physics Letters}\ }\textbf {\bibinfo {volume} {53}},\ \bibinfo {pages} {514} (\bibinfo {year} {1988})}\BibitemShut {NoStop}%
\bibitem [{\citenamefont {Nava}\ \emph {et~al.}(1986)\citenamefont {Nava}, \citenamefont {Bisi},\ and\ \citenamefont {Tu}}]{Nava1_1986}%
  \BibitemOpen
  \bibfield  {author} {\bibinfo {author} {\bibfnamefont {F.}~\bibnamefont {Nava}}, \bibinfo {author} {\bibfnamefont {O.}~\bibnamefont {Bisi}},\ and\ \bibinfo {author} {\bibfnamefont {K.~N.}\ \bibnamefont {Tu}},\ }\bibfield  {title} {\bibinfo {title} {Electrical transport properties of ${\mathrm{v}}_{3}$si, ${\mathrm{v}}_{5}$${\mathrm{si}}_{3}$, and ${\mathrm{vsi}}_{2}$ thin films},\ }\href {https://doi.org/10.1103/PhysRevB.34.6143} {\bibfield  {journal} {\bibinfo  {journal} {Phys. Rev. B}\ }\textbf {\bibinfo {volume} {34}},\ \bibinfo {pages} {6143} (\bibinfo {year} {1986})}\BibitemShut {NoStop}%
\bibitem [{\citenamefont {Clarke}(1968)}]{clarke1968}%
  \BibitemOpen
  \bibfield  {author} {\bibinfo {author} {\bibfnamefont {J.}~\bibnamefont {Clarke}},\ }\bibfield  {title} {\bibinfo {title} {The proximity effect between superconducting and normal thin films in zero field},\ }\href {https://doi.org/10.1051/jphyscol:1968201} {\bibfield  {journal} {\bibinfo  {journal} {Le Journal de Physique Colloques}\ }\textbf {\bibinfo {volume} {29}},\ \bibinfo {pages} {C2} (\bibinfo {year} {1968})}\BibitemShut {NoStop}%
\bibitem [{\citenamefont {Abrikosov}(2004)}]{abrikosov2004}%
  \BibitemOpen
  \bibfield  {author} {\bibinfo {author} {\bibfnamefont {A.~A.}\ \bibnamefont {Abrikosov}},\ }\bibfield  {title} {\bibinfo {title} {Nobel lecture: Type-ii superconductors and the vortex lattice},\ }\href {https://doi.org/10.1103/RevModPhys.76.975} {\bibfield  {journal} {\bibinfo  {journal} {Rev. Mod. Phys.}\ }\textbf {\bibinfo {volume} {76}},\ \bibinfo {pages} {975} (\bibinfo {year} {2004})}\BibitemShut {NoStop}%
\bibitem [{\citenamefont {Geim}\ \emph {et~al.}(1998)\citenamefont {Geim}, \citenamefont {Dubonos}, \citenamefont {Lok}, \citenamefont {Henini},\ and\ \citenamefont {Maan}}]{geim1998}%
  \BibitemOpen
  \bibfield  {author} {\bibinfo {author} {\bibfnamefont {A.}~\bibnamefont {Geim}}, \bibinfo {author} {\bibfnamefont {S.}~\bibnamefont {Dubonos}}, \bibinfo {author} {\bibfnamefont {J.}~\bibnamefont {Lok}}, \bibinfo {author} {\bibfnamefont {M.}~\bibnamefont {Henini}},\ and\ \bibinfo {author} {\bibfnamefont {J.}~\bibnamefont {Maan}},\ }\bibfield  {title} {\bibinfo {title} {Paramagnetic meissner effect in small superconductors},\ }\href {https://doi.org/https://doi.org/10.1038/24110} {\bibfield  {journal} {\bibinfo  {journal} {Nature}\ }\textbf {\bibinfo {volume} {396}},\ \bibinfo {pages} {144} (\bibinfo {year} {1998})}\BibitemShut {NoStop}%
\bibitem [{\citenamefont {Kosti\ifmmode~\acute{c}\else \'{c}\fi{}}\ \emph {et~al.}(1996)\citenamefont {Kosti\ifmmode~\acute{c}\else \'{c}\fi{}}, \citenamefont {Veal}, \citenamefont {Paulikas}, \citenamefont {Welp}, \citenamefont {Todt}, \citenamefont {Gu}, \citenamefont {Geiser}, \citenamefont {Williams}, \citenamefont {Carlson},\ and\ \citenamefont {Klemm}}]{kostic1996}%
  \BibitemOpen
  \bibfield  {author} {\bibinfo {author} {\bibfnamefont {P.}~\bibnamefont {Kosti\ifmmode~\acute{c}\else \'{c}\fi{}}}, \bibinfo {author} {\bibfnamefont {B.}~\bibnamefont {Veal}}, \bibinfo {author} {\bibfnamefont {A.~P.}\ \bibnamefont {Paulikas}}, \bibinfo {author} {\bibfnamefont {U.}~\bibnamefont {Welp}}, \bibinfo {author} {\bibfnamefont {V.~R.}\ \bibnamefont {Todt}}, \bibinfo {author} {\bibfnamefont {C.}~\bibnamefont {Gu}}, \bibinfo {author} {\bibfnamefont {U.}~\bibnamefont {Geiser}}, \bibinfo {author} {\bibfnamefont {J.~M.}\ \bibnamefont {Williams}}, \bibinfo {author} {\bibfnamefont {K.~D.}\ \bibnamefont {Carlson}},\ and\ \bibinfo {author} {\bibfnamefont {R.~A.}\ \bibnamefont {Klemm}},\ }\bibfield  {title} {\bibinfo {title} {Paramagnetic meissner effect in {Nb}},\ }\href {https://doi.org/10.1103/PhysRevB.53.791} {\bibfield  {journal} {\bibinfo  {journal} {Phys. Rev. B}\ }\textbf {\bibinfo {volume} {53}},\ \bibinfo {pages} {791} (\bibinfo {year} {1996})}\BibitemShut {NoStop}%
\bibitem [{\citenamefont {Vinokur}\ \emph {et~al.}(1990)\citenamefont {Vinokur}, \citenamefont {Kes},\ and\ \citenamefont {Koshelev}}]{vinokur1990}%
  \BibitemOpen
  \bibfield  {author} {\bibinfo {author} {\bibfnamefont {V.}~\bibnamefont {Vinokur}}, \bibinfo {author} {\bibfnamefont {P.}~\bibnamefont {Kes}},\ and\ \bibinfo {author} {\bibfnamefont {A.}~\bibnamefont {Koshelev}},\ }\bibfield  {title} {\bibinfo {title} {Flux pinning and creep in very anistropic high temperature superconductors},\ }\href {https://doi.org/https://doi.org/10.1016/0921-4534(90)90100-S} {\bibfield  {journal} {\bibinfo  {journal} {Physica C: Superconductivity}\ }\textbf {\bibinfo {volume} {168}},\ \bibinfo {pages} {29} (\bibinfo {year} {1990})}\BibitemShut {NoStop}%
\bibitem [{\citenamefont {Chaddah}(1991)}]{chaddah1991}%
  \BibitemOpen
  \bibfield  {author} {\bibinfo {author} {\bibfnamefont {P.}~\bibnamefont {Chaddah}},\ }\bibfield  {title} {\bibinfo {title} {Studies of irreversible magnetization in superconductors—a review},\ }\href {https://doi.org/https://doi.org/10.1007/BF02847211} {\bibfield  {journal} {\bibinfo  {journal} {Pramana J. Phys}\ }\textbf {\bibinfo {volume} {36}},\ \bibinfo {pages} {353} (\bibinfo {year} {1991})}\BibitemShut {NoStop}%
\bibitem [{\citenamefont {Karki}\ \emph {et~al.}(2011)\citenamefont {Karki}, \citenamefont {Xiong}, \citenamefont {Haldolaarachchige}, \citenamefont {Stadler}, \citenamefont {Vekhter}, \citenamefont {Adams}, \citenamefont {Young}, \citenamefont {Phelan},\ and\ \citenamefont {Chan}}]{2011_A.B.Karki}%
  \BibitemOpen
  \bibfield  {author} {\bibinfo {author} {\bibfnamefont {A.~B.}\ \bibnamefont {Karki}}, \bibinfo {author} {\bibfnamefont {Y.~M.}\ \bibnamefont {Xiong}}, \bibinfo {author} {\bibfnamefont {N.}~\bibnamefont {Haldolaarachchige}}, \bibinfo {author} {\bibfnamefont {S.}~\bibnamefont {Stadler}}, \bibinfo {author} {\bibfnamefont {I.}~\bibnamefont {Vekhter}}, \bibinfo {author} {\bibfnamefont {P.~W.}\ \bibnamefont {Adams}}, \bibinfo {author} {\bibfnamefont {D.}~\bibnamefont {Young}}, \bibinfo {author} {\bibfnamefont {W.}~\bibnamefont {Phelan}},\ and\ \bibinfo {author} {\bibfnamefont {J.~Y.}\ \bibnamefont {Chan}},\ }\bibfield  {title} {\bibinfo {title} {Physical properties of the noncentrosymmetric superconductor {Nb}$_{0.18}${Re}$_{0.82}$},\ }\href {https://doi.org/10.1103/PhysRevB.83.144525} {\bibfield  {journal} {\bibinfo  {journal} {Physical Review B}\ }\textbf {\bibinfo {volume} {83}},\ \bibinfo {pages} {144525} (\bibinfo {year} {2011})}\BibitemShut {NoStop}%
\bibitem [{\citenamefont {Xia}\ \emph {et~al.}(2022)\citenamefont {Xia}, \citenamefont {Zeng}, \citenamefont {Tian}, \citenamefont {Chen},\ and\ \citenamefont {Zaanen}}]{2022_Chuan-Yin_Xia}%
  \BibitemOpen
  \bibfield  {author} {\bibinfo {author} {\bibfnamefont {C.-Y.}\ \bibnamefont {Xia}}, \bibinfo {author} {\bibfnamefont {H.-B.}\ \bibnamefont {Zeng}}, \bibinfo {author} {\bibfnamefont {Y.}~\bibnamefont {Tian}}, \bibinfo {author} {\bibfnamefont {C.-M.}\ \bibnamefont {Chen}},\ and\ \bibinfo {author} {\bibfnamefont {J.}~\bibnamefont {Zaanen}},\ }\bibfield  {title} {\bibinfo {title} {Holographic abrikosov lattice: Vortex matter from black hole},\ }\href {https://doi.org/10.1103/PhysRevD.105.L021901} {\bibfield  {journal} {\bibinfo  {journal} {Phys. Rev. D}\ }\textbf {\bibinfo {volume} {105}},\ \bibinfo {pages} {L021901} (\bibinfo {year} {2022})}\BibitemShut {NoStop}%
\bibitem [{\citenamefont {Li}\ \emph {et~al.}(2015)\citenamefont {Li}, \citenamefont {Garcia}, \citenamefont {Franco}, \citenamefont {Lu}, \citenamefont {Lu}, \citenamefont {Rong}, \citenamefont {Shafiq}, \citenamefont {Chen}, \citenamefont {Liu}, \citenamefont {Liu}, \citenamefont {Song}, \citenamefont {Wei}, \citenamefont {Johnson}, \citenamefont {Luo},\ and\ \citenamefont {Feng}}]{2015_Y.Li}%
  \BibitemOpen
  \bibfield  {author} {\bibinfo {author} {\bibfnamefont {Y.}~\bibnamefont {Li}}, \bibinfo {author} {\bibfnamefont {J.}~\bibnamefont {Garcia}}, \bibinfo {author} {\bibfnamefont {G.}~\bibnamefont {Franco}}, \bibinfo {author} {\bibfnamefont {J.}~\bibnamefont {Lu}}, \bibinfo {author} {\bibfnamefont {K.}~\bibnamefont {Lu}}, \bibinfo {author} {\bibfnamefont {B.}~\bibnamefont {Rong}}, \bibinfo {author} {\bibfnamefont {B.}~\bibnamefont {Shafiq}}, \bibinfo {author} {\bibfnamefont {N.}~\bibnamefont {Chen}}, \bibinfo {author} {\bibfnamefont {Y.}~\bibnamefont {Liu}}, \bibinfo {author} {\bibfnamefont {L.}~\bibnamefont {Liu}}, \bibinfo {author} {\bibfnamefont {B.}~\bibnamefont {Song}}, \bibinfo {author} {\bibfnamefont {Y.}~\bibnamefont {Wei}}, \bibinfo {author} {\bibfnamefont {S.~S.}\ \bibnamefont {Johnson}}, \bibinfo {author} {\bibfnamefont {Z.}~\bibnamefont {Luo}},\ and\ \bibinfo {author} {\bibfnamefont {Z.}~\bibnamefont {Feng}},\ }\bibfield  {title} {\bibinfo {title} {{Critical magnetic fields of superconducting
  aluminum-substituted Ba$_8$Si$_{42}$Al$_4$ clathrate}},\ }\href {https://doi.org/10.1063/1.4921702} {\bibfield  {journal} {\bibinfo  {journal} {Journal of Applied Physics}\ }\textbf {\bibinfo {volume} {117}},\ \bibinfo {pages} {213912} (\bibinfo {year} {2015})}\BibitemShut {NoStop}%
\bibitem [{\citenamefont {Matsuda}\ and\ \citenamefont {Shimahara}(2007)}]{2007_Y.Matsuda}%
  \BibitemOpen
  \bibfield  {author} {\bibinfo {author} {\bibfnamefont {Y.}~\bibnamefont {Matsuda}}\ and\ \bibinfo {author} {\bibfnamefont {H.}~\bibnamefont {Shimahara}},\ }\bibfield  {title} {\bibinfo {title} {Fulde--ferrell--larkin--ovchinnikov state in heavy fermion superconductors},\ }\href {https://doi.org/https://doi.org/10.1143/JPSJ.76.051005} {\bibfield  {journal} {\bibinfo  {journal} {Journal of the Physical Society of Japan}\ }\textbf {\bibinfo {volume} {76}},\ \bibinfo {pages} {051005} (\bibinfo {year} {2007})}\BibitemShut {NoStop}%
\bibitem [{\citenamefont {Wu}\ \emph {et~al.}(1994)\citenamefont {Wu}, \citenamefont {Keren}, \citenamefont {Le}, \citenamefont {Luke}, \citenamefont {Sternlieb}, \citenamefont {Uemura}, \citenamefont {Johnston}, \citenamefont {Cho},\ and\ \citenamefont {Gehring}}]{1994_W.D.Wu}%
  \BibitemOpen
  \bibfield  {author} {\bibinfo {author} {\bibfnamefont {W.}~\bibnamefont {Wu}}, \bibinfo {author} {\bibfnamefont {A.}~\bibnamefont {Keren}}, \bibinfo {author} {\bibfnamefont {L.}~\bibnamefont {Le}}, \bibinfo {author} {\bibfnamefont {G.}~\bibnamefont {Luke}}, \bibinfo {author} {\bibfnamefont {B.}~\bibnamefont {Sternlieb}}, \bibinfo {author} {\bibfnamefont {Y.}~\bibnamefont {Uemura}}, \bibinfo {author} {\bibfnamefont {D.}~\bibnamefont {Johnston}}, \bibinfo {author} {\bibfnamefont {B.}~\bibnamefont {Cho}},\ and\ \bibinfo {author} {\bibfnamefont {P.}~\bibnamefont {Gehring}},\ }\bibfield  {title} {\bibinfo {title} {Magnetic penetration depth in $\textnormal{V}_3\textnormal{Si}$ and $\textnormal{Li}\textnormal{Ti}_2\textnormal{O}_4$ measured by $\mu${SR}},\ }\href {https://doi.org/https://doi.org/10.1007/BF02068956} {\bibfield  {journal} {\bibinfo  {journal} {Hyperfine Interactions}\ }\textbf {\bibinfo {volume} {86}},\ \bibinfo {pages} {615} (\bibinfo {year} {1994})}\BibitemShut {NoStop}%
\bibitem [{\citenamefont {Yethiraj}\ \emph {et~al.}(1999)\citenamefont {Yethiraj}, \citenamefont {Christen}, \citenamefont {Paul}, \citenamefont {Miranovic},\ and\ \citenamefont {Thompson}}]{1999_M.Yethiraj}%
  \BibitemOpen
  \bibfield  {author} {\bibinfo {author} {\bibfnamefont {M.}~\bibnamefont {Yethiraj}}, \bibinfo {author} {\bibfnamefont {D.}~\bibnamefont {Christen}}, \bibinfo {author} {\bibfnamefont {D.~M.}\ \bibnamefont {Paul}}, \bibinfo {author} {\bibfnamefont {P.}~\bibnamefont {Miranovic}},\ and\ \bibinfo {author} {\bibfnamefont {J.}~\bibnamefont {Thompson}},\ }\bibfield  {title} {\bibinfo {title} {Flux lattice symmetry in $\textnormal{V}_3\textnormal{Si}$: Nonlocal effects in a high-$\kappa$ superconductor},\ }\href {https://doi.org/10.1103/PhysRevLett.82.5112} {\bibfield  {journal} {\bibinfo  {journal} {Physical review letters}\ }\textbf {\bibinfo {volume} {82}},\ \bibinfo {pages} {5112} (\bibinfo {year} {1999})}\BibitemShut {NoStop}%
\bibitem [{\citenamefont {Ding}\ \emph {et~al.}(2023)\citenamefont {Ding}, \citenamefont {Zhao}, \citenamefont {Jiang}, \citenamefont {Wang}, \citenamefont {Feng},\ and\ \citenamefont {Zhang}}]{2023_Shuyue.Ding}%
  \BibitemOpen
  \bibfield  {author} {\bibinfo {author} {\bibfnamefont {S.}~\bibnamefont {Ding}}, \bibinfo {author} {\bibfnamefont {D.}~\bibnamefont {Zhao}}, \bibinfo {author} {\bibfnamefont {T.}~\bibnamefont {Jiang}}, \bibinfo {author} {\bibfnamefont {H.}~\bibnamefont {Wang}}, \bibinfo {author} {\bibfnamefont {D.}~\bibnamefont {Feng}},\ and\ \bibinfo {author} {\bibfnamefont {T.}~\bibnamefont {Zhang}},\ }\bibfield  {title} {\bibinfo {title} {Surface structure and multigap superconductivity of $\textnormal{V}_3\textnormal{Si}$ (111) revealed by scanning tunneling microscopy},\ }\href {https://doi.org/https://doi.org/10.1007/s44214-023-00028-y} {\bibfield  {journal} {\bibinfo  {journal} {Quantum Frontiers}\ }\textbf {\bibinfo {volume} {2}},\ \bibinfo {pages} {3} (\bibinfo {year} {2023})}\BibitemShut {NoStop}%
\bibitem [{\citenamefont {Christen}\ \emph {et~al.}(1985)\citenamefont {Christen}, \citenamefont {Kerchner}, \citenamefont {Sekula},\ and\ \citenamefont {Chang}}]{christen1985}%
  \BibitemOpen
  \bibfield  {author} {\bibinfo {author} {\bibfnamefont {D.}~\bibnamefont {Christen}}, \bibinfo {author} {\bibfnamefont {H.}~\bibnamefont {Kerchner}}, \bibinfo {author} {\bibfnamefont {S.}~\bibnamefont {Sekula}},\ and\ \bibinfo {author} {\bibfnamefont {Y.}~\bibnamefont {Chang}},\ }\bibfield  {title} {\bibinfo {title} {Flux-line lattice anisotropy in v3si: Observation and interpretation},\ }\href {https://doi.org/https://doi.org/10.1016/0378-4363(85)90509-1} {\bibfield  {journal} {\bibinfo  {journal} {Physica B+C}\ }\textbf {\bibinfo {volume} {135}},\ \bibinfo {pages} {369} (\bibinfo {year} {1985})}\BibitemShut {NoStop}%
\bibitem [{\citenamefont {Moshchalkov}\ \emph {et~al.}(1996)\citenamefont {Moshchalkov}, \citenamefont {Baert}, \citenamefont {Metlushko}, \citenamefont {Rosseel}, \citenamefont {Van~Bael}, \citenamefont {Temst}, \citenamefont {Jonckheere},\ and\ \citenamefont {Bruynseraede}}]{moshchalkov1996}%
  \BibitemOpen
  \bibfield  {author} {\bibinfo {author} {\bibfnamefont {V.}~\bibnamefont {Moshchalkov}}, \bibinfo {author} {\bibfnamefont {M.}~\bibnamefont {Baert}}, \bibinfo {author} {\bibfnamefont {V.}~\bibnamefont {Metlushko}}, \bibinfo {author} {\bibfnamefont {E.}~\bibnamefont {Rosseel}}, \bibinfo {author} {\bibfnamefont {M.}~\bibnamefont {Van~Bael}}, \bibinfo {author} {\bibfnamefont {K.}~\bibnamefont {Temst}}, \bibinfo {author} {\bibfnamefont {R.}~\bibnamefont {Jonckheere}},\ and\ \bibinfo {author} {\bibfnamefont {Y.}~\bibnamefont {Bruynseraede}},\ }\bibfield  {title} {\bibinfo {title} {Magnetization of multiple-quanta vortex lattices},\ }\href {https://doi.org/10.1103/PhysRevB.54.7385} {\bibfield  {journal} {\bibinfo  {journal} {Physical Review B}\ }\textbf {\bibinfo {volume} {54}},\ \bibinfo {pages} {7385} (\bibinfo {year} {1996})}\BibitemShut {NoStop}%
\bibitem [{\citenamefont {Maniv}\ \emph {et~al.}(2001)\citenamefont {Maniv}, \citenamefont {Zhuravlev}, \citenamefont {Vagner},\ and\ \citenamefont {Wyder}}]{maniv2001}%
  \BibitemOpen
  \bibfield  {author} {\bibinfo {author} {\bibfnamefont {T.}~\bibnamefont {Maniv}}, \bibinfo {author} {\bibfnamefont {V.}~\bibnamefont {Zhuravlev}}, \bibinfo {author} {\bibfnamefont {I.}~\bibnamefont {Vagner}},\ and\ \bibinfo {author} {\bibfnamefont {P.}~\bibnamefont {Wyder}},\ }\bibfield  {title} {\bibinfo {title} {Vortex states and quantum magnetic oscillations in conventional type-{II} superconductors},\ }\href {https://doi.org/10.1103/RevModPhys.73.867} {\bibfield  {journal} {\bibinfo  {journal} {Reviews of Modern Physics}\ }\textbf {\bibinfo {volume} {73}},\ \bibinfo {pages} {867} (\bibinfo {year} {2001})}\BibitemShut {NoStop}%
\bibitem [{\citenamefont {Campbell}\ and\ \citenamefont {Evetts}(1972)}]{campbell1972}%
  \BibitemOpen
  \bibfield  {author} {\bibinfo {author} {\bibfnamefont {A.}~\bibnamefont {Campbell}}\ and\ \bibinfo {author} {\bibfnamefont {J.}~\bibnamefont {Evetts}},\ }\bibfield  {title} {\bibinfo {title} {Flux vortices and transport currents in type ii superconductors},\ }\href {https://doi.org/10.1080/00018737200101288} {\bibfield  {journal} {\bibinfo  {journal} {Advances in Physics}\ }\textbf {\bibinfo {volume} {21}},\ \bibinfo {pages} {199} (\bibinfo {year} {1972})}\BibitemShut {NoStop}%
\bibitem [{\citenamefont {Sosolik}\ \emph {et~al.}(2003)\citenamefont {Sosolik}, \citenamefont {Stroscio}, \citenamefont {Stiles}, \citenamefont {Hudson}, \citenamefont {Blankenship}, \citenamefont {Fein},\ and\ \citenamefont {Celotta}}]{Sosolik2003}%
  \BibitemOpen
  \bibfield  {author} {\bibinfo {author} {\bibfnamefont {C.~E.}\ \bibnamefont {Sosolik}}, \bibinfo {author} {\bibfnamefont {J.~A.}\ \bibnamefont {Stroscio}}, \bibinfo {author} {\bibfnamefont {M.~D.}\ \bibnamefont {Stiles}}, \bibinfo {author} {\bibfnamefont {E.~W.}\ \bibnamefont {Hudson}}, \bibinfo {author} {\bibfnamefont {S.~R.}\ \bibnamefont {Blankenship}}, \bibinfo {author} {\bibfnamefont {A.~P.}\ \bibnamefont {Fein}},\ and\ \bibinfo {author} {\bibfnamefont {R.~J.}\ \bibnamefont {Celotta}},\ }\bibfield  {title} {\bibinfo {title} {Real-space imaging of structural transitions in the vortex lattice of ${\mathrm{v}}_{3}\mathrm{Si}$},\ }\href {https://doi.org/10.1103/PhysRevB.68.140503} {\bibfield  {journal} {\bibinfo  {journal} {Phys. Rev. B}\ }\textbf {\bibinfo {volume} {68}},\ \bibinfo {pages} {140503} (\bibinfo {year} {2003})}\BibitemShut {NoStop}%
\bibitem [{\citenamefont {Rosenstein}\ and\ \citenamefont {Li}(2010)}]{rosenstein2010}%
  \BibitemOpen
  \bibfield  {author} {\bibinfo {author} {\bibfnamefont {B.}~\bibnamefont {Rosenstein}}\ and\ \bibinfo {author} {\bibfnamefont {D.}~\bibnamefont {Li}},\ }\bibfield  {title} {\bibinfo {title} {Ginzburg-landau theory of type {II} superconductors in magnetic field},\ }\href {https://doi.org/10.1103/RevModPhys.82.109} {\bibfield  {journal} {\bibinfo  {journal} {Reviews of modern physics}\ }\textbf {\bibinfo {volume} {82}},\ \bibinfo {pages} {109} (\bibinfo {year} {2010})}\BibitemShut {NoStop}%
\bibitem [{\citenamefont {De~Gennes}(2018)}]{de2018}%
  \BibitemOpen
  \bibfield  {author} {\bibinfo {author} {\bibfnamefont {P.-G.}\ \bibnamefont {De~Gennes}},\ }\href {https://doi.org/10.1201/9780429497032} {\emph {\bibinfo {title} {Superconductivity of metals and alloys}}}\ (\bibinfo  {publisher} {CRC press},\ \bibinfo {year} {2018})\BibitemShut {NoStop}%
\bibitem [{\citenamefont {Golubov}\ and\ \citenamefont {Hartmann}(1994)}]{1994_A.A.Golubov}%
  \BibitemOpen
  \bibfield  {author} {\bibinfo {author} {\bibfnamefont {A.}~\bibnamefont {Golubov}}\ and\ \bibinfo {author} {\bibfnamefont {U.}~\bibnamefont {Hartmann}},\ }\bibfield  {title} {\bibinfo {title} {Electronic structure of the abrikosov vortex core in arbitrary magnetic fields},\ }\href {https://doi.org/10.1103/PhysRevLett.72.3602} {\bibfield  {journal} {\bibinfo  {journal} {Physical review letters}\ }\textbf {\bibinfo {volume} {72}},\ \bibinfo {pages} {3602} (\bibinfo {year} {1994})}\BibitemShut {NoStop}%
\bibitem [{\citenamefont {Boaknin}\ \emph {et~al.}(2003)\citenamefont {Boaknin}, \citenamefont {Tanatar}, \citenamefont {Paglione}, \citenamefont {Hawthorn}, \citenamefont {Ronning}, \citenamefont {Hill}, \citenamefont {Sutherland}, \citenamefont {Taillefer}, \citenamefont {Sonier}, \citenamefont {Hayden} \emph {et~al.}}]{2003_Etienne_Boaknin}%
  \BibitemOpen
  \bibfield  {author} {\bibinfo {author} {\bibfnamefont {E.}~\bibnamefont {Boaknin}}, \bibinfo {author} {\bibfnamefont {M.}~\bibnamefont {Tanatar}}, \bibinfo {author} {\bibfnamefont {J.}~\bibnamefont {Paglione}}, \bibinfo {author} {\bibfnamefont {D.}~\bibnamefont {Hawthorn}}, \bibinfo {author} {\bibfnamefont {F.}~\bibnamefont {Ronning}}, \bibinfo {author} {\bibfnamefont {R.}~\bibnamefont {Hill}}, \bibinfo {author} {\bibfnamefont {M.}~\bibnamefont {Sutherland}}, \bibinfo {author} {\bibfnamefont {L.}~\bibnamefont {Taillefer}}, \bibinfo {author} {\bibfnamefont {J.}~\bibnamefont {Sonier}}, \bibinfo {author} {\bibfnamefont {S.}~\bibnamefont {Hayden}}, \emph {et~al.},\ }\bibfield  {title} {\bibinfo {title} {Heat conduction in the vortex state of {NbSe}$_2$: Evidence for multiband superconductivity},\ }\href {https://doi.org/10.1103/PhysRevLett.90.117003} {\bibfield  {journal} {\bibinfo  {journal} {Physical review letters}\ }\textbf {\bibinfo {volume} {90}},\ \bibinfo {pages} {117003} (\bibinfo {year}
  {2003})}\BibitemShut {NoStop}%
\end{thebibliography}%
\end{document}